\documentclass[12pt]{iopart}
\usepackage{harvard}

\expandafter\let\csname equation*\endcsname\relax

\expandafter\let\csname endequation*\endcsname\relax

\usepackage{multirow}
\usepackage{amsmath}
\usepackage{graphicx,xcolor}
\usepackage{subcaption}

\usepackage{iopams}  
\begin{document}
\citationmode{abbr}

\title[TB Cough Classification]{Automatic Cough Classification for Tuberculosis Screening in a Real-World Environment}

\author{Madhurananda Pahar$^1$, Marisa Klopper$^2$, Byron Reeve$^2$, Rob Warren$^2$, Grant Theron$^2$, Thomas Niesler$^1$}

\address{$^1$Department of Electrical and Electronic Engineering, Stellenbosch University, South Africa \\
	$^2$SAMRC Centre for Tuberculosis Research, Division of Molecular Biology and Human Genetics, DSI/NRF Centre of Excellence for Biomedical Tuberculosis Research, Faculty of Medicine and Health Sciences, Stellenbosch University, South Africa}
%	\{mpahar, marisat, byronreeve, gtheron, rw1, trn\}@sun.ac.za}
\ead{\{mpahar, marisat, byronreeve, rw1, gtheron, trn\}@sun.ac.za}
\vspace{10pt}
\begin{indented}
	\item[]August 2021
\end{indented}

\begin{abstract}
	\\
	\textit{Objective:}
	%	We present first results showing that it is possible to 
	The automatic discrimination between the coughing sounds produced by patients with tuberculosis (TB) and those produced by patients with other lung ailments. 
	\textit{Approach:}
	We present experiments based on a dataset of 1358 forced cough recordings obtained in a developing-world clinic from 16 patients with confirmed active pulmonary TB and 35 patients suffering from respiratory conditions suggestive of TB but confirmed to be TB negative. 
	Using nested cross-validation, we have trained and evaluated five machine learning classifiers: logistic regression (LR), support vector machines (SVM), k-nearest neighbour (KNN), multilayer perceptrons (MLP) and convolutional neural networks (CNN). 
	\textit{Main Results:}
	Although classification is possible in all cases, the best performance is achieved using LR. 
	In combination with feature selection by sequential forward selection (SFS), our best LR system achieves an area under the ROC curve (AUC) of 0.94 using 23 features selected from a set of 78 high-resolution mel-frequency cepstral coefficients (MFCCs). 
	This system achieves a sensitivity of 93\% at a specificity of 95\% 
	and thus exceeds the 90\% sensitivity at 70\% specificity specification considered by the World Health Organisation (WHO) as a minimal requirement for a community-based TB triage test. 
	\textit{Significance:}	
	The automatic classification of cough audio sounds, when applied to symptomatic patients requiring investigation for TB, can meet the WHO triage specifications for the identification of patients who should undergo expensive molecular downstream testing.
	This makes it a promising and viable means of low cost, easily deployable frontline screening for TB, which can benefit especially developing countries with a heavy TB burden. 
\end{abstract}

%\vspace{2pc}

\section{Introduction}

%\subsection{What is TB?}
Tuberculosis (TB) is a bacterial infection primarily of the lungs and globally a leading cause of death~\cite{floyd2018global}. 
Modern diagnostic tests rely on costly laboratory procedures requiring specialized equipment~\cite{dewan2006feasibility,bwanga2009direct,konstantinos2010testing,global2017practical}.
However, TB is generally most prevalent in low-income settings and is responsible for 95\% of deaths due to infectious disease in developing countries~\cite{TB-WHO}. 
Due to a high index of suspicion in such high-incidence settings, these expensive tests are frequently conducted on patients who meet symptom criteria for TB investigation but cough due to other lung ailments. 
In fact, most people investigated for TB do not suffer from the disease~\cite{chang2008chronic}. 

The simplest form of TB triaging relies on self-reported symptoms.
Although there is no cost involved, this has low specificity, resulting in over-testing. 
Furthermore, TB is also associated with stigmatisation, which may result in under-reporting of symptoms, leading to under-testing and consequent inadequate care~\cite{nathavitharana2019guidance}. 
Thus, there is a need for a low-cost, point-of-care screening test, such as the automatic classification of cough sounds, which would allow a more efficient and widespread application of molecular testing. 
If such an objective audio-based test, which is “specimen-free", were accurate enough, it might offer an improvement in the standard of care~\cite{naidoo2017south}. 
%This test we are evaluating would be objective and more accurate, offering a massive improvement over the current TB triage standard of care. 
%Thus, there is a need for a low-cost, point-of-care screening test, such as the automatic classification of cough sounds. 

%\subsection{What is a cough?} 
Coughing is a common symptom of respiratory disease and caused by an explosive expulsion of air from the airways \cite{simonsson1967role}. 
However, the effect of coughing on the respiratory system is known to vary~\cite{higenbottam2002chronic}. 
For example, lung diseases can cause the airway to be either restricted or obstructed and this can influence the cough acoustics~\cite{chung2008prevalence}.
It has also been postulated that the glottis behaves differently under different pathological conditions and that this makes it possible to distinguish between coughs due to asthma, bronchitis and pertussis (whooping cough)~\cite{korpavs1996analysis}. 
Therefore, the automatic classification of the acoustic signals associated with coughing in order to detect lung diseases like TB seems to be a reasonable avenue of investigation. 

%\subsection{Can we really classify coughs?}

Vocal audio has been used in various disease classification studies, including the recent COVID-19 pandemic \cite{hassan2020covid,brown2020exploring}. 
%However, it has been found that coughs are more capable to classify respiratory disease \cite{pahar2021covidbreath}. 
Aspects of speech such as phonation and vowel sounds have been used to detect Parkinson’s disease by applying machine learning \cite{almeida2019detecting,hemmerling2019parkinson}. 
Respiratory disease such as asthma bronchiale (AB) has also been successfully detected by analysing cough frequency \cite{marsden2016objective}.
Finally, cough sounds have been used in the diagnosis and screening of pulmonary diseases such as AB, chronic obstructive pulmonary disease (COPD) and TB \cite{infante2017use}. 
%The automatic classification of coughing has been considered by a number of authors in the past.
The voluntary coughs produced by AB and COPD patients were successfully distinguished from those produced by healthy participants using discriminant analysis with an accuracy of between~85\% and~90\% in~\cite{knocikova2008wavelet}. 
The detection of coughing associated with asthma was considered in~\cite{al2013signal}, while pertussis was detected with good accuracy using logistic regression (LR) in~\cite{pramono2016cough}.
The early detection of congestive heart failure (CHF) and COPD, which can increase the fatality rate in an ageing population, using a random forest classifier was considered in~\cite{windmon2018tussiswatch}. 
%In this case a random forest classifier was shown to achieve good performance.
%has been used to classify if a cough series indicates disease and the classifier has achieved 80\% accuracy on a small dataset containing only 36 subjects. 
Successful detection of the seal-like barking cough that can occur in children who suffer from croup or laryngotracheobronchitis has been reported in~\cite{sharan2018automatic}.
%on a dataset containing 479 patients. The sensitivity and specificity has been achieved as high as 92.31\% and 85.29\% respectively. 
More recently, domain-specific features like mel-frequency cepstral coefficients (MFCCs) and zero crossing rate (ZCR) have shown promise when used to classify coughs~\cite{rudraraju2020cough} due to pneumonia \cite{sotoudeh2020artificial} and COVID-19 \cite{pahar2020covid,pahar2021covidbreath}. 
The recently introduced generative adversarial neural network architecture has also proven to be successful in respiratory disease classification \cite{ramesh2020coughgan}. 

This work is a direct extension of our previous work in which we demonstrated that it is possible to discriminate between the coughs of TB sufferers and healthy controls using logistic regression~\cite{botha2018detection}. 
However, studies that involve only cases with a condition and healthy controls have well-known limitations \cite{rutjes2005case}. 
We now show that it is also possible to distinguish between the coughs of TB patients and the coughs of patients suffering from other lung ailments and for whom TB was excluded as a diagnosis.
In contrast to the controlled environment in which our previous recordings were made, the audio data we use in this work were collected at a TB clinic and include substantial environmental noise, thereby directly addressing the practical scenario encountered at primary health facilities in developing countries. Patients presenting to these clinics are typically ill and it is necessary to establish the likelihood of TB disease for further referral.
This referral is typically achieved by collecting an infectious and difficult to handle specimen (sputum) which is tested using an expensive (in a developing world context) test that requires laboratory expertise and specialised equipment. 
As in our previous work, we focus on automatic cough classification, and assume that the detection of the start and end of coughing has been reliably achieved. 
We acknowledge that, by sidestepping this detection step, difficult practical challenges, such as the processing of cough spasms, have been left for future work.

To perform our experiments, it was necessary to compile a new corpus of coughing sounds, gathered from patients who suffer from symptoms of TB but do not necessarily have TB.
This is a priority population for the World Health Organisation (WHO) TB triage test target product profile. 
Our new corpus is of a similar extent to that used in our previous work, but is compiled in a more realistic environment, representative of the developing-world primary health care environment in which TB screening would likely to be performed.
We therefore believe it provides a first direct affirmation that automatic sound analysis is a promising and viable approach to TB screening in high-incidence settings.

The structure of the remainder of this paper is as follows.
The next section will describe our new dataset and how it was compiled.
Sections \ref{sec:feature-extraction} and \ref{sec:classifiers} describes the acoustic features we consider and the machine learning techniques we evaluate. 
Section~\ref{sec:training} presents our method of parameter training and hyperparameter optimisation, followed by the experimental results in Section~\ref{sec:results}. 
Results are discussed in Section \ref{sec:discussion} and finally Section~\ref{sec:conclusion} concludes the paper.

\section{Data Collection}

\subsection{Collection Setup}
\begin{figure}
	\centerline{\includegraphics[width=0.9\textwidth]{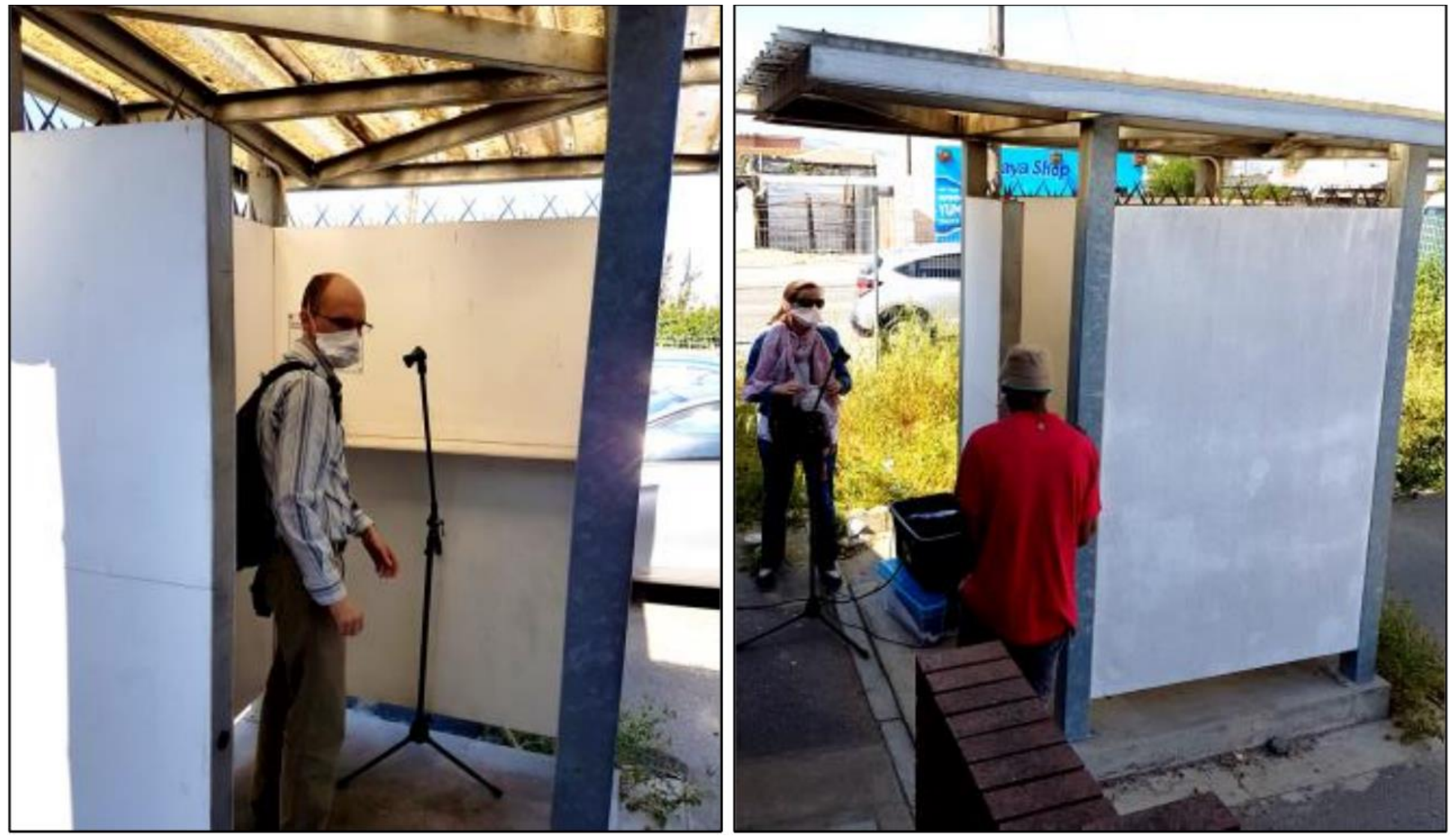}}
	\caption{\textbf{Recording booth}: Cough recordings were made in a standard cross-ventilated sputum collection booth situated outside on the premises of a primary health care clinic in a high-density urban neighbourhood of Cape Town, South Africa. The first panel shows the inside of the booth while the second shows a member of the research staff explaining the recording process to a study participant. The recording environment can be considered challenging due to a constant and considerable level of environmental noise. Co-authors Dr. Byron Reeve (left panel) and Dr. Marisa Klopper (right panel) gave their consent to have their visible faces appear in this figure. }
	\label{fig:recording-booth}
\end{figure}

Our data were collected in a busy primary health care clinic in Cape Town, South Africa, where mobile recording equipment was deployed in an outside cross-ventilated sputum collection booth, as shown in Figure \ref{fig:recording-booth}. 
This setup is representative of the typical real-world clinic environment in a developing country in which low-cost, easily-deployable automatic TB screening is most urgently needed.
All recordings were taken between 10 am and 4 pm by two health care workers without a technical background but who were trained to operate the recording equipment.
The recording area was not fitted with additional acoustic protection and was exposed to noise from a consistently high number of patients and staff attending the clinic while the surrounding streets were busy with pedestrians, pets and vehicles. 
Thus, our dataset contains a considerable amount of environmental noise, and is representative of the scenario in which a TB screening test would likely be deployed.  
No attempts were made to de-noise the data since we were specifically interested in the performance that can be expected in a real-world scenario. 
Furthermore, experience in the related field of automatic speech recognition has shown that noise reduction performed before model training is often not beneficial and may reduce robustness to varying input conditions~\cite{caballero2018machine}. 
This study was approved by the Faculty of Health Sciences Research Ethics Committee of Stellenbosch University (N14/10/136) and the City of Cape Town (10483). Informed consent was granted for all patients in this study. 

\begin{figure}
	\centerline{\includegraphics[width=0.8\textwidth]{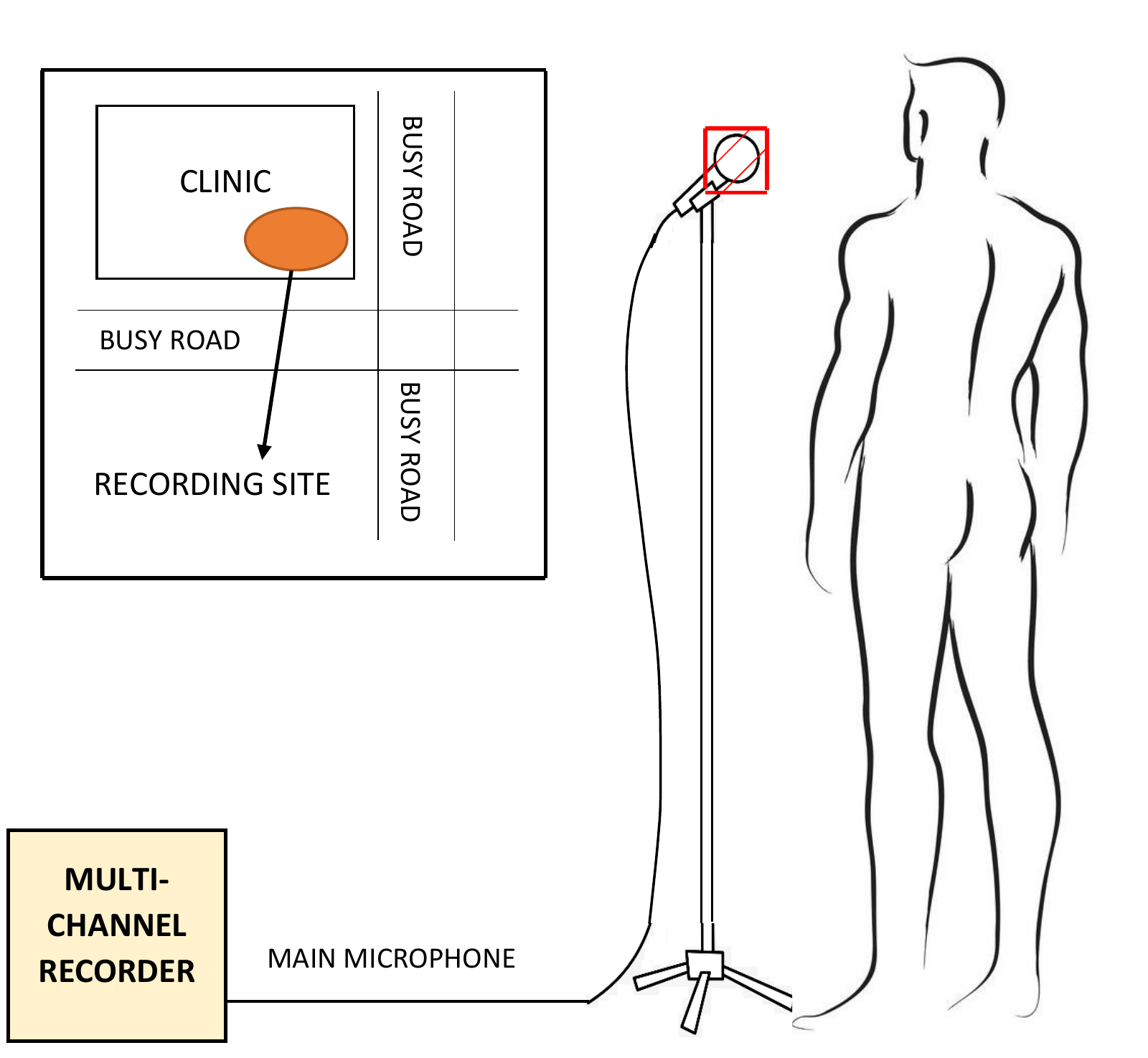}}
	\caption{\textbf{Recording setup:} During recording, the patient stands in front of a microphone covered by a standard N95 mask at a distance of approximately 10 to 15 centimetres. For each patient, the recording session lasted approximately 5 minutes on average. Data collection took place next to a busy road, as shown in the inset. }
	\label{fig:recording-setUp}
\end{figure}

\subsection{Recording Setup and Annotation}
A ZOOM F8N field recorder was used to record the audio captured by a R{\O}DE M3 condenser microphone \cite{hsu1998high,todorovic2015multilayer}, 
%A ZOOM F8N MultiTrack Field Recorder was used to record three channels: the main microphone, the left stethoscope and the right stethoscope. 
%A Rode M3 condenser microphone 
covered by a standard N95 mask which was replaced after each patient. 
%with two protective layers to assist with infection control, one of which was replaced after each patient. 
%Condenser microphones are directional and thus able to focus on speaker's voice and this way we have captured minimum environmental noise, recording the best-possible audio quality with minimum signal-to-noise ratio (SNR) in a real-world environment \cite{hsu1998high,todorovic2015multilayer}. 
Informal listening tests indicated that the mask did not substantially affect the quality of the recorded audio signal. 
The health care workers ensured that the gap maintained between the patient and the microphone was 10 to 15 centimetres (Figure \ref{fig:recording-setUp}).  
Each patient was prompted to count from one to ten, cough, take a few deep breaths and then cough again, thus producing at least two bursts of coughs. 
All patients in our study were suffering from some sort of respiratory disease. 
So, when they were prompted to cough, they produced a bout of voluntary coughs due to the irritation in their respiratory system \cite{simonsson1967role}. 
In a real-world diagnostic scenario, the patient would be asked to produce a voluntary cough. 
Therefore, our methods are appropriate to best approximate the practical application of the TB classifier.

\begin{figure}
	\centerline{\includegraphics[width=0.9\textwidth]{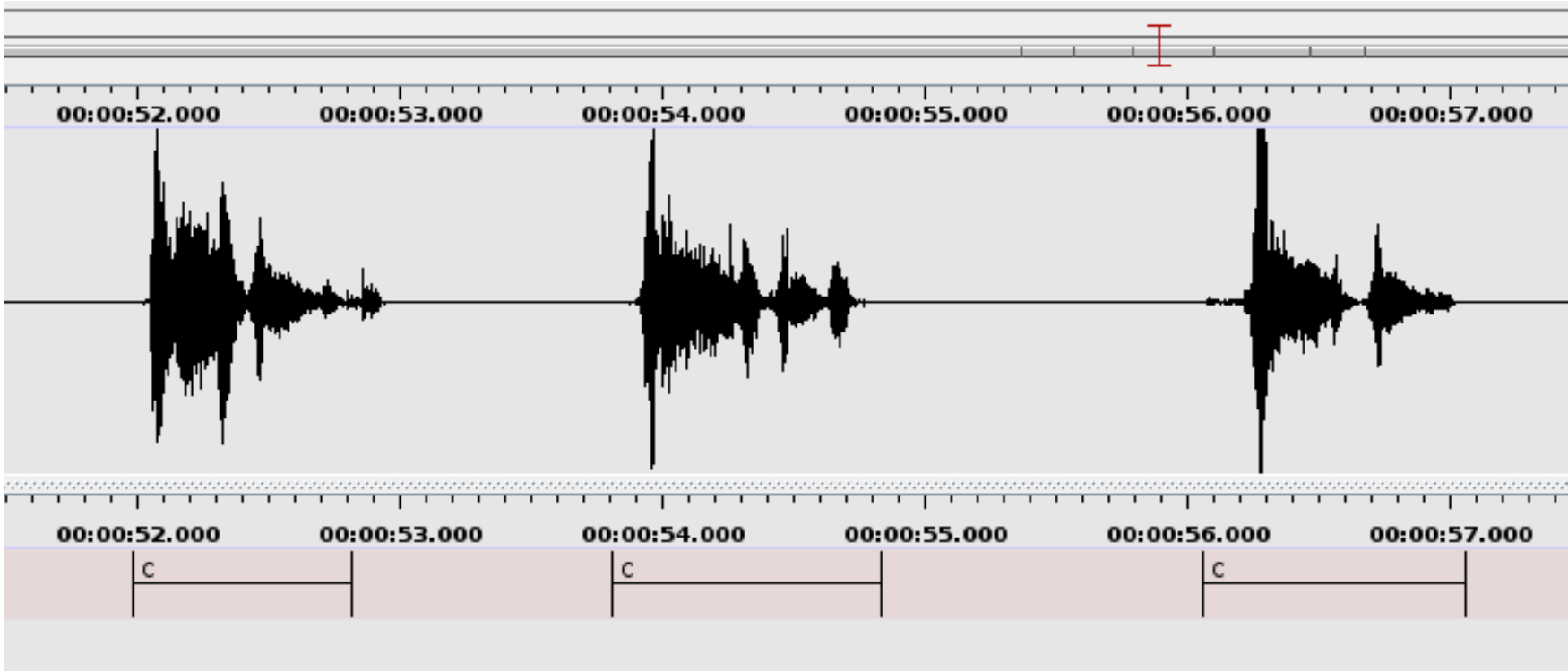}}
	\caption{\textbf{Cough annotation process:} The start and end times of all cough events were manually annotated by the label `c' in the audio recording using the ELAN software. This figure shows the manual annotation of three consecutive cough events. The three labelled cough events include two, three and two individual cough onsets respectively. The manual annotation process indicated the start and the end times of audio containing coughing, but did not label such onsets. However, it revealed that the average number of such onsets is almost equal across the TB and non-TB classes. Each manually annotated cough event was subsequently automatically divided into sections, as described in Section \ref{subsec:segments}. 
		%		The first event shows two onsets of the cough event, the second one shows three such onsets and final event two onsets. This manual annotation has enabled us to extract only the cough events from the entire audio recording. 
	}
	\label{fig:ELAN-annot}
\end{figure}

All audio recordings were sampled at 44.1 kHz.
The portions of the resulting audio recordings that contain coughing were manually annotated using the ELAN multimedia software~\cite{wittenburg2006elan} as shown in Figure~\ref{fig:ELAN-annot}. 
We note that these manually annotated stretches of voluntary coughing often contain several cough onsets.
The number of these onsets for all 1358 cough events (Table \ref{table:Dataset-Summary}) is 3124, indicating an average of 2.3 onsets per cough event. 
Among the 402 TB and 956 non-TB cough events, there are 973 and 2151 cough onsets respectively, indicating 2.42$\pm$0.83 onsets per TB cough event and 2.25$\pm$0.91 onsets per non-TB cough event. These means and standard deviations suggest that the number of cough bursts or onsets per cough event is not an influential factor in the TB cough classification task.
In this research, we make no attempt to automatically identify the boundaries of such onset subdivisions, and this remains an aspect of our ongoing work.
We will refer to these stretches of audio containing coughing, such as those annotated in Figure~\ref{fig:ELAN-annot}, as \textit{cough events} in the remainder of this paper.

%\begin{figure}
%	\centerline{\includegraphics[width=0.9\textwidth]{images/ELAN-annot}}
%	\caption{\textbf{Cough annotation process:} The start and end times of all cough events were manually annotated by the label `c' in the audio recording using the ELAN software. This figure shows the manual annotation of three consecutive cough events. The first event shows two onsets of the cough event, the second one shows three such onsets and final event one onset. This manual annotation has enabled us to extract only the cough events from the entire audio recording. }
%	\label{fig:ELAN-annot}
%\end{figure}

\subsection{Dataset Description}
\begin{figure}
	\centerline{\includegraphics[width=0.9\textwidth]{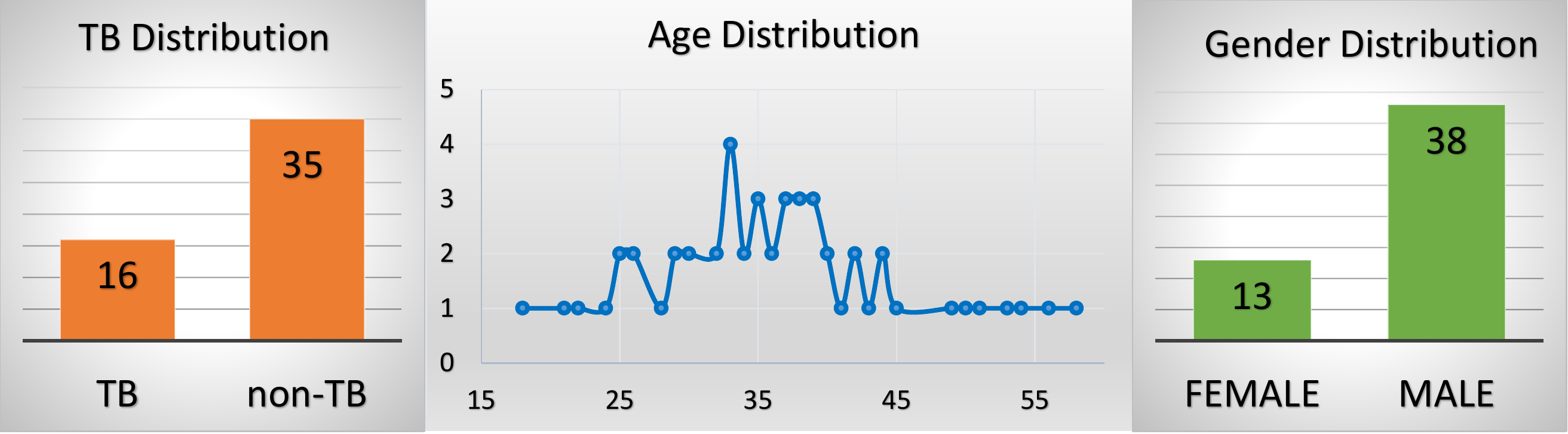}}
	\caption{\textbf{Demographic distribution of the patients} show that most of the participants are middle-aged and male patients dominant over the female patients.}
	\label{fig:TB-dataset}
\end{figure}

Our dataset currently contains coughs from 16 TB and 35 non-TB patients and
most participants are male with an average age of 38 
%and female participants appear almost evenly in our dataset as there are 6 TB female and 7 non-TB female patients 
(Figure \ref{fig:TB-dataset}).

All participants were TB suspects that self-reported an involuntary cough suggestive of an underlying lung pathology.
Clinical work was limited to bacteriological TB diagnosis for the purpose of the study. 
The participants were not seen by a medical doctor, but rather audio samples were collected by the health care workers.
Differential diagnosis for diseases other than TB was impractical to collect since they would, in general, require several additional tests, and even then the diagnosis is often based on treatment-related symptom resolution.
Therefore, patients were only tested for TB by standardised methods for the purpose of the study and no alternative diagnoses were established apart from TB. 
The inclusion and exclusion criteria for the participants are listed in Table~\ref{table:inclusion-exclusion}.
This information was collected during a formal interview conducted by the health care workers.

\begin{table}[h]
	%	\small
	%	\footnotesize
	\caption{\textbf{Inclusion and exclusion criteria} for participants who are included in our dataset summarised in Table \ref{table:Dataset-Summary}.} % title name of the table
	\centering % centering table
	\begin{center}
		\begin{tabular}{ p{7.5cm} | p{7.5cm} }
			\hline
			\hline
			\\[-0.5em]
			\textbf{Inclusion} & \textbf{Exclusion} \\
			\\[-0.5em]
			\hline
			\hline
			\\[-2em]
			\begin{itemize}
				\item Age $>$ 18 years, AND
				
				\item if HIV negative: 
				\begin{itemize}
					\item Cough for $>$ 2 weeks, AND
					\item Additional symptoms (any of night sweats, fever, weight loss, coughing blood) 
				\end{itemize}
				
				\item if HIV positive: 
				\begin{itemize}
					\item Cough for any duration, AND
					\begin{itemize}
						\item Additional symptoms (any of night sweats, fever, weight loss, coughing blood), OR
						\item In regular contact with known TB case 
					\end{itemize}
					
				\end{itemize}
				
			\end{itemize} 
			
			&
			
			\begin{itemize}
				\item No consent 
				\item On TB treatment during the 60 days prior to enrolment 
				\item Unable to provide sputum specimens for testing to confirm TB 
				\item Unable to provide cough audio 
			\end{itemize} \\

			\hline
			\hline
		\end{tabular}
	\end{center}
	\label{table:inclusion-exclusion}
\end{table}

Table~\ref{table:Dataset-Summary} describes the dataset and shows that there is an imbalance between the number of TB and non-TB coughs. 
We have used AUC as the performance measure as it has a higher degree of discriminancy than some other existing performance measures such as accuracy for imbalanced datasets \cite{rakotomamonjy2004optimizing,huang2005using,fawcett2006introduction}. 
%This imbalance has been addressed by using
%%To compensate for this imbalance, we have used 
%AUC as the method of classifier evaluation, since this metric is insensitive to skewed datasets ~\cite{fawcett2006introduction}. 
The length of all coughs in our dataset is 1045 seconds (17.42 minutes). 
TB coughs are on average 0.74 seconds long with a standard deviation of 0.31, while non-TB coughs are on average 0.78 seconds long with a standard deviation of 0.39. 
Therefore, we note that coughs produced by TB patients are of comparable length to the coughs produced by suffers from other lung ailments. 
This is in contrast to our previous finding that the coughs produced by TB patients are both longer and greater in number than those produced by healthy individuals~\cite{botha2018detection}. 

\begin{table}[h]
	\small
	%	\footnotesize
	\setlength{\tabcolsep}{4pt} % Default value: 6pt
	\caption{\textbf{Dataset description}: Composition of the dataset used for experimentation.
		All recorded patients were ill, either with tuberculosis (TB) or with a different lung ailment.} % title name of the table
	\centering % centering table
	\begin{center}
		\begin{tabular}{ l | c c c c c c }
			\hline
			\hline
			& \textbf{No. of} & \textbf{No. of} & \textbf{Avg coughs} & \textbf{Avg length} & \textbf{Avg SNR} & \textbf{Total length}\\
			& \textbf{patients} & \textbf{coughs} & \textbf{per patient} & \textbf{of coughs} & \textbf{of coughs} & \textbf{of coughs} \\
			\hline
			TB             & 16  &  402 &  25.1  & 0.74 sec & 33.27$\pm$15.11 dB & 299 sec \\
			\hline
			Non-TB         & 35  &  956 &  27.3     & 0.78 sec & 33.93$\pm$19.24 dB  & 746 sec \\
			
			\hline
			\hline
			\textbf{Total} & 51  & 1358 &  26.6  & 0.77 sec & 33.72 dB  &  1045 sec \\
			
			\hline
			\hline
		\end{tabular}
	\end{center}
	\label{table:Dataset-Summary}
\end{table}

All recordings were carried out inside the same recording booth (Figures \ref{fig:recording-booth} and \ref{fig:recording-setUp}) and there was no link between the time or date of recording and whether the subject was TB positive or TB negative. 
Hence, we can assume that the SNR is independent of the TB status. 
This was confirmed by informal listening checks applied to the
audio recordings as well as 
%by SNR estimates for both the TB and the non-TB coughs. 
%The environmental noise present in the recordings therefore cannot influence the TB classification task. 
by calculating SNR estimates
%~\cite{johnson2006signal,fgee1999comparing} 
for the recordings of TB and non-TB coughs,  as listed in Table~\ref{table:Dataset-Summary}.
The table shows that the average SNR is 33.27 dB and 33.93 dB for TB and non-TB coughs respectively, and that the difference between these figures ($0.67$ dB) is much smaller than the standard deviation of the SNR estimates of both classes. 
%Environmental noises such as chattering, running engine, moving vehicle weren't present in all the coughs. 
%However, we have estimated the SNR for all the coughs and listed in Table \ref{table:Dataset-Summary}. 
%It shows that TB coughs on average contain SNR of 33.27 dB and non-TB coughs contain a very similar SNR of 33.93 dB. 
%The standard deviation of the SNR present in the TB coughs is 15.11 dB and for non-TB coughs, 19.24 dB. 
We have used the Equation \ref{eq:SNR} to estimate SNR \cite{johnson2006signal,fgee1999comparing}. 

\begin{equation}
	\text{SNR(dB)} = 10 \log \frac{P_s}{P_n}
	\label{eq:SNR}
\end{equation}
where, $P_s$ is the signal power of the cough audio and $P_n$ is the signal power of the background noise i.e. the entire audio recording except the coughs, breaths and spoken digits uttered by the patients. 

\begin{figure}
	\centering
	\begin{subfigure}{.5\textwidth}
		\centering
		\includegraphics[width=\linewidth]{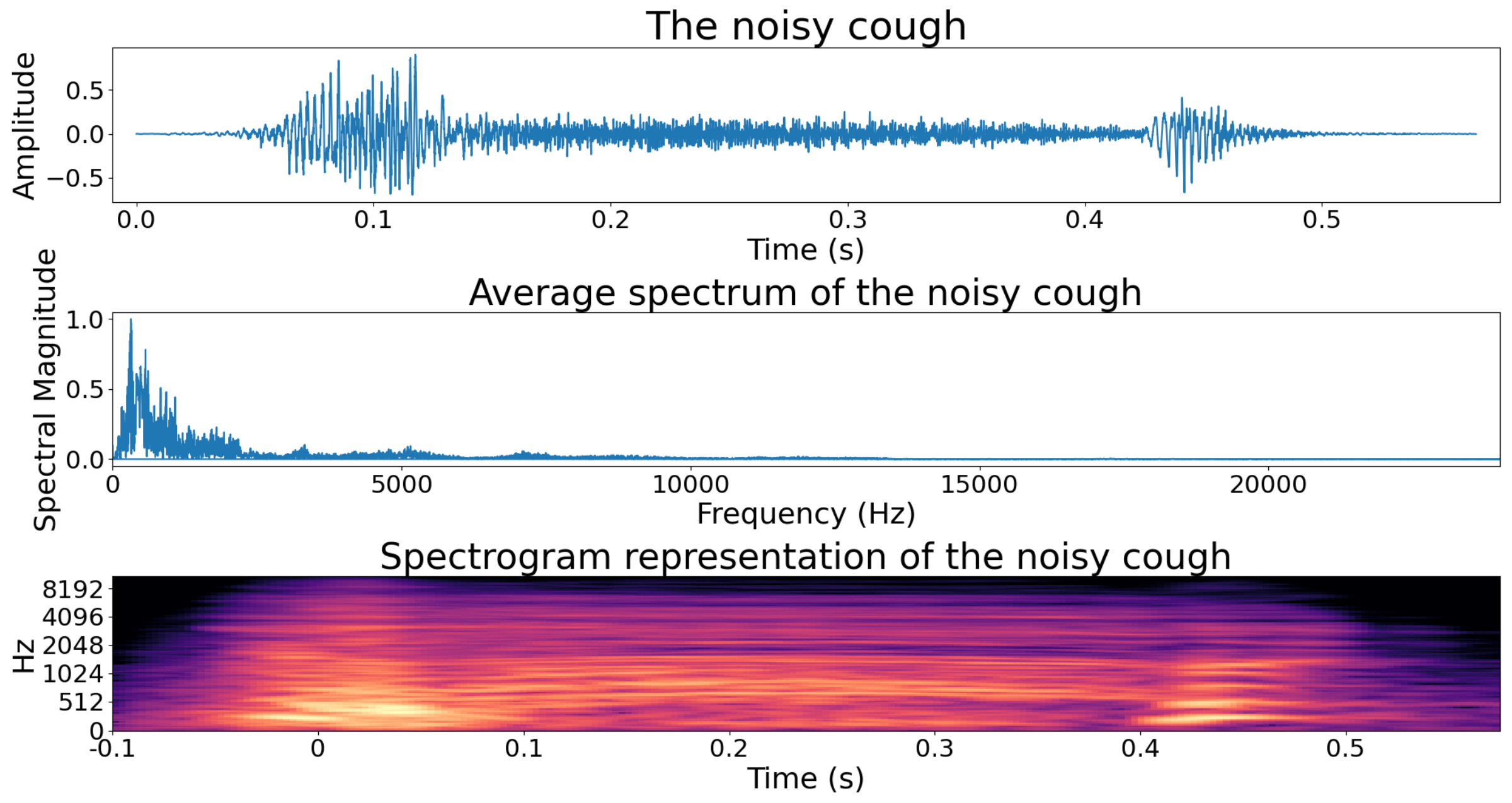}
		\caption{A noisy cough of SNR $\approx$ 22 dB.}
		\label{fig:sub1}
	\end{subfigure}%
	\begin{subfigure}{.5\textwidth}
		\centering
		\includegraphics[width=\linewidth]{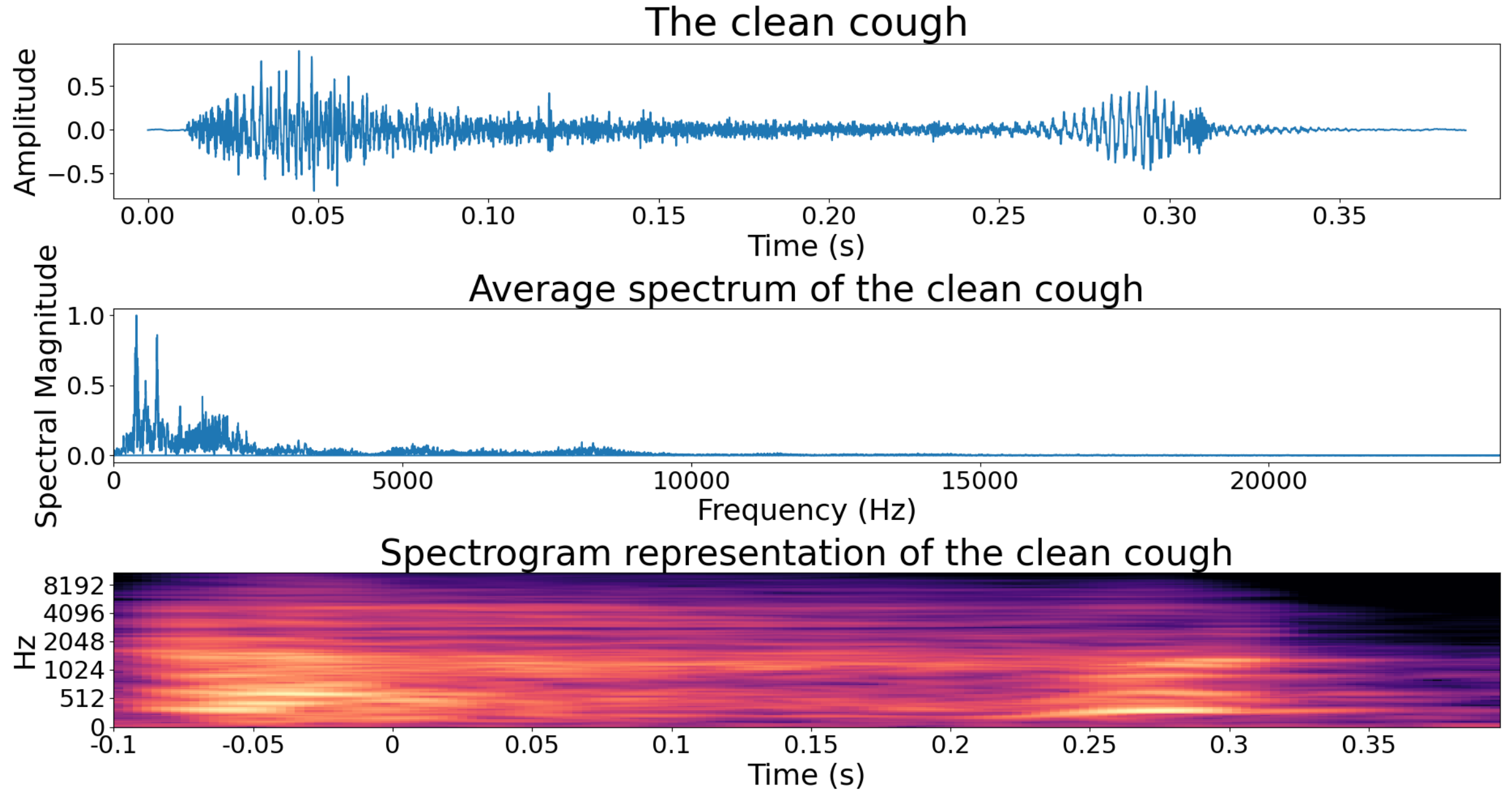}
		\caption{A clean cough of SNR $\approx$ 45 dB.}
		\label{fig:sub2}
	\end{subfigure}
	\caption{\textbf{Examples of a noisy and a clean cough:} The noisy cough contains human chattering between its two phases and the clean cough contains minimal background noise. }
	\label{fig:cough-compare}
\end{figure}

For illustration, we show two coughs in Figure~\ref{fig:cough-compare}. 
The first with human chattering between its two onsets (SNR $\approx$ 22dB) and the second with little background noise (SNR $\approx$ 45dB).

\section{Feature Extraction}\label{sec:feature-extraction}
\begin{figure}
	\centerline{\includegraphics[width=0.9\textwidth]{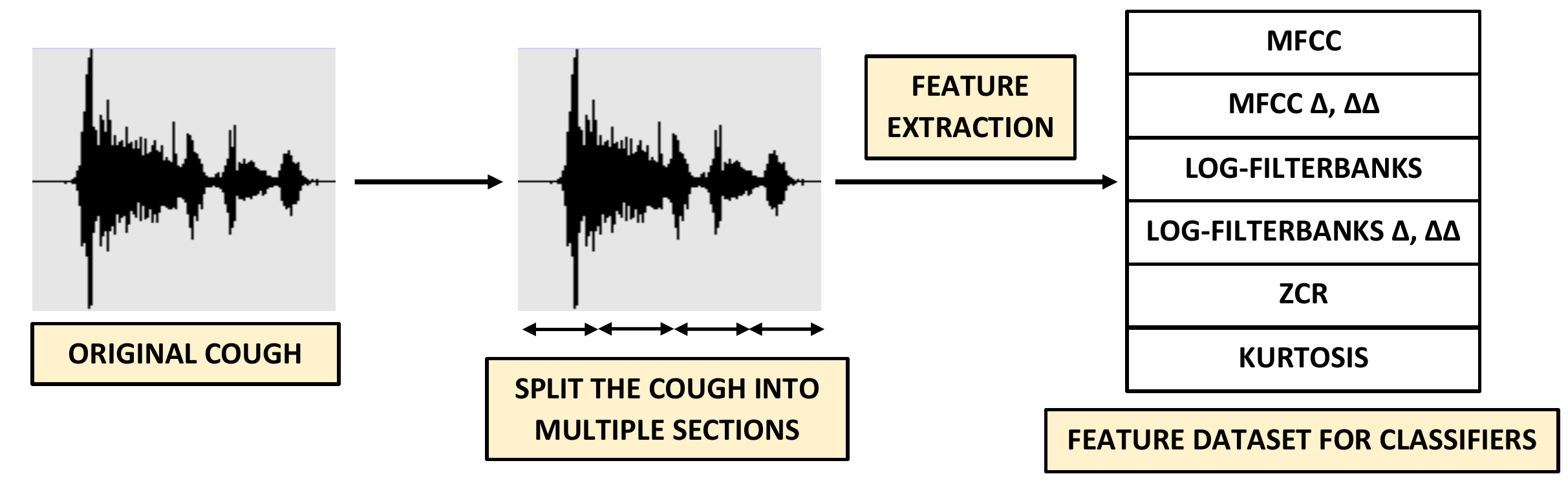}}
	\caption{\textbf{Feature extraction:} Each raw cough recording, as shown in Figure~\ref{fig:ELAN-annot}, is automatically split into individual sections after which features including MFCCs (including velocity $\Delta$ and acceleration $\Delta \Delta$), linearly spaced log-filterbank energies (including velocity $\Delta$ and acceleration $\Delta \Delta$), zero crossing rate and kurtosis are extracted. For example, when using 13 MFCCs, (13 $\times$ 3 + 2) = 41 features including $\Delta$, $\Delta \Delta$, zero crossing rate and kurtosis are extracted for each section. The number of sections, MFCCs and linearly spaced filters are used as feature extraction hyperparameters listed in Table \ref{table:feat-hyper-parameter}. }
	\label{fig:feat-extract}
\end{figure}

The feature extraction process is illustrated in Figure~\ref{fig:feat-extract}. 
No filtering or pre-processing has been applied to the cough audio.
%; rather raw audio has been used for feature extraction. 
We have considered mel-frequency cepstral coefficients (MFCCs), log-filterbank energies, zero-crossing rate (ZCR) and kurtosis as features.

%\begin{figure}
%	\centerline{\includegraphics[width=0.9\textwidth]{images/feat_extract}}
%	\caption{\textbf{Feature extraction:} Each raw cough recording, as shown in Figure~\ref{fig:ELAN-annot}, is automatically segmented into individual sections after which features including MFCCs (including velocity $\Delta$ and acceleration $\Delta \Delta$), linearly spaced log-filterbank energies (including velocity $\Delta$ and acceleration $\Delta \Delta$), zero crossing rate and kurtosis are extracted. For example, when using 13 MFCCs, (13 $\times$ 3 + 2) = 41 features including $\Delta$, $\Delta \Delta$, zero crossing rate and kurtosis are extracted for each segment. The number of segments, MFCCs and linearly spaced filters are used as feature extraction hyperparameters listed in Table \ref{table:feat-hyper-parameter}. }
%	\label{fig:feat-extract}
%\end{figure}

\subsection{MFCCs}

Mel-frequency cepstral coefficients (MFCCs) have been used very successfully as features in audio analysis and especially in automatic speech recognition~\cite{WeiHan2006,pahar_coding_2020}. 
They have also been found to be useful for differentiating dry coughs from wet coughs \cite{chatrzarrin2011feature}. 
MFCCs were computed using the following standard procedure: 

\begin{itemize}
	\item The audio signal is divided into short frames and the fast Fourier transform (FFT) is computed for each. 
	
	\begin{equation}
		f_{mel}(f) = 2595 \times (1+\frac{f}{700})
		\label{eq:mel-filters}
	\end{equation}
	
	\item Mel-scaled filterbanks are computed (Equation \ref{eq:mel-filters}) and the log-power spectrum is calculated. 
	%	\item The discrete Fourier transformation (DFT) is applied to the output of the mel-filterbanks (Equation \ref{eq:DCT}) and a certain number $k$ of the resulting coefficients are retained. $n$ ranges over the frame length and $i$, number of frames. $S_{i}(k)$ is the complex DFT, where $i$ denotes the frame number corresponding to the time-domain frame and $ 1\leq k \leq N$ ($N$ is the length of the DFT). 
	\item The discrete cosine transformation (DCT) is applied to the output of the mel-filterbanks and a certain number of the resulting coefficients are retained. 
	\item The long-term mean of each coefficient is calculated and then subtracted. 
	\item Inter-frame derivatives (velocity $\Delta$) and second-order derivatives (acceleration $\Delta \Delta$) are computed for each coefficient and appended to the already computed MFCCs~\cite{azmy2017feature}. Equation~\ref{eq:delta-delta} shows the computation of the delta coefficient $d_{t}$ for the frames $c_{t-n}$ to $c_{t+n}$ and the number of samples ($N$) is 2.
\end{itemize}

%\begin{equation}
%S_{i}(k) = \sum\limits_{j=1}^P m_j . cos( \frac{\pi_i}{P} (j - 0.5) )
%\label{eq:DCT}
%\end{equation}

%\begin{equation}
%S_{i}(k) = \sum\limits_{n=1}^N s_i(n) h(n) e^{-j2\pi kn/N}
%\label{eq:DCT}
%\end{equation}

\begin{equation}
	d_{t} = \frac{\sum\limits_{n=1}^N n(c_{t+n} - c_{t-n})}{2 \sum\limits_{n=1}^N n^2}
	\label{eq:delta-delta}
\end{equation}

\subsection{Log-Filterbank Energies}

These features \cite{garreton2011telephone} consist of the log energies computed after applying $F$ linearly spaced overlapping triangular filters to the frame power spectrum $\mathfrak{S}(t)$ which is computed using Equation~\ref{eq:log-power}, where $X(t)$ is the FFT of the audio frame and $N$ is the number of samples. 

\begin{equation}
	\mathfrak{S}(t) = \frac{1}{N} |X(t)|^2
	\label{eq:log-power}
\end{equation}

\subsection{ZCR}

The zero-crossing-rate (ZCR) is the number of times the signal changes sign within a frame, as indicated in Equation~\ref{eq:ZCR} \cite{bachu2010voiced}. 
ZCR indicates the variability present in the signal.

\begin{equation}
	ZCR = \frac{1}{T - 1} \sum\limits_{t=1}^{T-1} \lambda (x(t) x(t-1) < 0)
	\label{eq:ZCR}
\end{equation}

In Equation~\ref{eq:ZCR}, $\lambda = 1$ when the sign of $x(t)$ and $x(t-1)$ differ and $\lambda = 0$ when the sign of $x(t)$ and $x(t-1)$ is the same and $T$ is the frame length in samples. 

\subsection{Kurtosis}
The kurtosis indicates the tailedness of a probability density \cite{decarlo1997meaning} and specially the prevalence of higher amplitudes in an audio signal. 
Kurtosis has been calculated according to Equation~\ref{eq:kurt}, where $\mu$ is the mean and $\sigma$ is the standard deviation. 

\begin{equation}
	\Lambda_x = \frac{1}{T - 1} \sum\limits_{t=1}^{T-1} \frac{x(t) - \mu}{\sigma^4}
	\label{eq:kurt}
\end{equation}

\subsection{Feature extraction hyperparameters}\label{subsec:segments}

The feature extraction process is influenced by a number of hyperparamamters, listed in Table~\ref{table:feat-hyper-parameter}. 
Each cough event is first divided evenly into between 1 and 4 sections.
From each section, non-overlapping consecutive frames are used to extract features which are averaged. 
Finally, these average feature vectors for each section are concatenated, increasing the dimensionality of the feature vector by a factor equal to the number of sections. 
The number of sections is a feature extraction hyperparameter, indicated in Table \ref{table:feat-hyper-parameter}. 
For example, when using 13 MFCCs; (3 $\times$ 13 + 2) = 41 features are extracted from each frame making the dimension of the feature vector (41 $\times$ 1). 
Therefore when four sections are used, the feature vector presented to the classifier has dimensions (164 $\times$ 1).

\begin{table}[h]
	\caption{\textbf{Feature extraction hyperparameters} for which the results are shown in Table \ref{table:classifier-results}.} % title name of the table
	\centering % centering table
	\begin{center}
		\begin{tabular}{ c | c | c }
			\hline
			\hline
			\textbf{Hyperparameter} & \textbf{Description} & \textbf{Range} \\
			\hline
			\hline
			Frame length & Length of the frames (in samples) & $2^k$ where \\
			($\mathcal{F} =$) & from which features were extracted &  $k=8, 9, ..., 12$\\
			\hline
			No. of sections & Number of sections into & \multirow{2}{*}{1, 2, 3, 4} \\
			($\mathcal{S} =$) & which frames were grouped &  \\
			\hline
			No. of linearly spaced filters & Number of filters used to & 40 to 200 \\
			($\mathcal{B} =$) & extract log-filerbank energies & in steps of 20 \\
			\hline
			No. of MFCCs & Number of lower order MFCCs  & \multirow{2}{*}{13, 26, 39}\\
			($\mathcal{M}=$) & coefficients retained &  \\ 
			\hline
			\hline
		\end{tabular}
	\end{center}
	\label{table:feat-hyper-parameter}
\end{table}

For log-filterbank energies, the number of filters in the filterbank is another hyperparameter.
For MFCCs, the number of coefficients that are computed can also be varied.
While 13 MFCCs are generally accepted to reflect the level of discrimination of the human auditory system, we have also considered a larger number of MFCCs in our experiments.

%The hyperparameters used for the feature extraction are listed in Table~\ref{table:feat-hyper-parameter}. 
The frame length corresponds to the number of time-domain samples per frame. 
Since the audio was sampled at 44.1 kHz, by varying the frame lengths from 256 to 4096 samples, features are extracted from frame durations varying between approximately 5 and 100 msec.
By varying the number of log-filterbank filters and MFCCs, the spectral resolution of the features was varied.
% and the features were not normalised before classifier training. 
The only form of feature normalisation applied before classifier training was cepstral mean normalisation which was performed on a per-recording basis.

\section{Classifier Description}\label{sec:classifiers}
Five classifiers have been considered for discrimination between TB and non-TB coughs.  

\subsection{Logistic Regression (LR)}
Logistic regression (LR) models have in some clinical situations been found to outperform more sophisticated classifiers~\cite{christodoulou2019systematic}. 
We came to the same conclusion in our previous work into TB cough classification~\cite{botha2018detection}, where these models comfortably outperformed hidden-Markov model (HMM) and decision tree (DT) classifiers. 
%Researchers have used LR since 19th century and ridge estimators in LR has been used since 1992 \cite{le1992ridge}. 

The output of an LR model varies between 0 and 1, making it very useful in binary classification. 
It can also be considered as a single neuron neural network.
The output of an LR classifier is given by:

\begin{equation}
	P = \frac{1}{1 + e^{-(a+\mathbf{b}\mathbf{x})}} 
	\label{eq:LR}
\end{equation}
where, the scalar $a$ and the vector $\mathbf{b}$ are the parameters of the model and $P$ is the classifier probability. 

We have considered the gradient descent weight regularisation strength $\nu_1$ as well as lasso ($l1$ penalty) and ridge ($l2$ penalty) estimators to be hyperparameters which were optimised during nested k-fold cross-validation (Figure \ref{fig:nested-k-fold}). 
%We note  = gradient descent weight regularisation strength, $\nu_2$ = $l1$ penalty and $\nu_3$ = $l2$ penalty. 
We have used Equation~\ref{eq:l1-penalty} to estimate the $l1$ penalty ($\nu_2$) and Equation~\ref{eq:l2-penalty} to estimate $l2$ penalty ($\nu_3$) while optimising the loss function of the LR model \cite{yamashita2003interior,tsuruoka2009stochastic}. 

\begin{equation}
	\nu_2 = \sum\limits_{i=1}^{n} (\mathbf{y}_i - \sum\limits_{j} \mathbf{x}_{ij} \beta_{j})^2 + \lambda \sum\limits_{j=1}^{p} |\beta_{j}|
	\label{eq:l1-penalty}
\end{equation}

\begin{equation}
	\nu_3 = \sum\limits_{i=1}^{n} (\mathbf{y}_i - \sum\limits_{j} \mathbf{x}_{ij} \beta_{j})^2 + \lambda \sum\limits_{j=1}^{p} |\beta_{j}|^2
	\label{eq:l2-penalty}
\end{equation}

In Equations~\ref{eq:l1-penalty} and~\ref{eq:l2-penalty}, $\mathbf{x}$ and~$\mathbf{y}$ are  the independent and dependent variables respectively.

\subsection{k-Nearest Neighbour (KNN)}

The k-nearest neighbour classifier bases its decision on the class labels of the $k$ nearest neighbours in the training set. 
This machine learning algorithm has in the past successfully been able to both detect \cite{monge2018robust,pramono2019automatic,vhaduri2019nocturnal} and classify \cite{wang2006environmental,pramono2016cough,pahar2021covidbreath} sounds such as coughs and snores. 
Our KNN classifier uses the Euclidean distance to calculate similarity.

%This machine learning technique calculates the minimum Euclidean distance, which can be explained by Equation \ref{eq:knn} \cite{zhang2007ml}. 
%
%\begin{equation}
%%d(p, q) = d(q, p) = \sqrt{(q_1 - p_1)^2 + (q_2 - p_2)^2 + \dots + (q_k - p_k)^2 } 
%%= \sqrt{\sum_{i=1}^{n} (q_i - p_i)^2}
%\begin{split}
%d(\mathbf{p}, \mathbf{q}) = d(\mathbf{q}, \mathbf{p}) = \sqrt{(\mathbf{q}_1 - \mathbf{p}_1)^2 + (\mathbf{q}_2 - \mathbf{p}_2)^2 + \dots + (\mathbf{q}_k - \mathbf{p}_k)^2 } \\
%= \sqrt{\sum_{i=1}^{k} (\mathbf{q}_i - \mathbf{p}_i)^2}
%\end{split}
%\label{eq:knn}
%\end{equation}
%where $\mathbf{p}$ and $\mathbf{q}$ are two different feature vectors. 

\subsection{Support Vector Machines (SVM)}

Support vector machines (SVM) classifiers have performed well in both detecting and classifying coughing sounds~\cite{tracey2011cough,bhateja2019pre,sharan2017cough}.
We have used both linear and non-linear SVM classifiers, based on the computation in Equation \ref{eq:SVM-2}. 

%\begin{equation} \label{eq:SVM-1}
%g(\mathbf{x}) = \mathbf{w}^T\mathbf{x} + b
%\end{equation}
%where, $g(\mathbf{x})$ is related to $||\mathbf{w}||$ in such a way that, $g(\mathbf{x}) \geq 1$ and minimise $||\mathbf{w}||$. 
%
%Then, the SVM is calculated in Equation \ref{eq:SVM-2}. 

\begin{equation} \label{eq:SVM-2}
	\phi(\mathbf{w}) = \frac{1}{2} \mathbf{w}^T\mathbf{w} - J(\mathbf{w}, b, a)
\end{equation}
where, $\mathbf{w}$ is the weight vector, $a$ and $b$ are the coefficients and $J(\mathbf{w}, b, a)$ is the term to minimise during hyperparameter optimisation for the parameters listed in Table \ref{table:class-hyper-parameter}.

\subsection{Multilayer Perceptron (MLP)}

The LR model described in the previous section was intended to be our baseline, and we hoped to improve classification performance using a multilayer perceptron (MLP) neural network.
Unlike LR, the MLP is capable of learning non-linear relationships by using multiple layers of neurons to separate input and output. 
The MLP classifier is based on Equation~\ref{eq:MLP}, which shows the computation of a single neuron. 

\begin{equation}
	y = \phi ( \sum\limits_{i=1}^{n} w_i x_i + b) = \phi (\mathbf{w}^T\mathbf{x} + b)
	\label{eq:MLP}
\end{equation}

Here, $\mathbf{x}$ is the input-vector, $\mathbf{w}$ is the weight-vector, $b$ is the bias and $\phi$ is the non-linear activation function. 
The weights and the bias are optimised during supervised training.

We have optimised the loss function by using $l2$ penalty estimator, shown in Equation~\ref{eq:l2-penalty} and stochastic gradient descent. 
The $l2$ penalty estimator, stochastic gradient descent learning rate and the number of hidden layers have been considered as the hyperparameters (Table~\ref{table:class-hyper-parameter}) which were optimised using nested k-fold cross-validation (Figure~\ref{fig:nested-k-fold}).  
% This is described in the table
%We note $\xi_1$ = number of hidden layers, $\xi_2$ = $l2$ penalty estimator (Equation \ref{eq:l2-penalty}) and $\xi_3$ = stochastic gradient descent learning rate for MLP. 

\subsection{Convolutional Neural Network (CNN)}

A convolutional neural network (CNN) is a popular deep neural network architecture which has proved particularly effective in image classification~\cite{krizhevsky2017imagenet,lawrence1997face}. 
%For example, in the past two decades CNNs have been applied successfully to complex tasks such as face recognition~\cite{lawrence1997face}. 
The core of a CNN can be expressed by Equation~\ref{eq:CNN-equation}, where $net(t, f)$ is the output of the convolutional layer \cite{albawi2017understanding}.  

%For an image of size $N \times N$, with the filter size $F$ and the stride size $S$;  
%
%\begin{equation} \label{eq:CNN-stride}
%O = 1 + \frac{N-F}{S}
%\end{equation}
%
%Next, while applying the number of the layers of the zero-padding as $P$;
%
%\begin{equation} \label{eq:CNN-padding}
%O = 1 + \frac{N+2P-F}{S}
%\end{equation}
%
%Finally, the convolution output of the next layer $net(t, f)$ is calculated by following the Equation \ref{eq:CNN-equation}.

%The final convolution output $net(t, f)$ is calculated by following the Equation \ref{eq:CNN-equation}.

\begin{equation} \label{eq:CNN-equation}
	net(t, f) = (\mathbf{x} \ast \mathbf{w})[t, f] = \sum_{m} \sum_{n} \mathbf{x}[m, n] \mathbf{w}[t-m, f-n]
\end{equation}
In this equation, $\ast$ is the convolution operation, $\mathbf{w}$ is the filter or kernel matrix and $\mathbf{x}$ is the input image. 
The rectified linear activation function was used in the hidden layers and the softmax activation function is applied in the final layer of our CNN architecture \cite{qi2017comparison}.
%Finally an activation function is applied before final output. We have used softmax function \cite{qi2017comparison} and it can be expressed in Equation \ref{eq:CNN-softmax}. 
%
%\begin{equation} \label{eq:CNN-softmax}
%%\sigma(z)_i = \frac{e^{z_i}}{\sum_{j=1}^{K} e^{z_i}} \; \; \; \; \; \; \; \; \forall i = 1,
%\sigma(z)_i = \frac{e^{z_i}}{\sum_{j=1}^{K} e^{z_i}} \; \; \; \; \forall i = 1, \cdots, K \text{ and } z=(z_1, \cdots, z_K) \in \mathbb{R}^K
%\end{equation}
%where, $K$ is the size of the $net(t, f)$ in Equation \ref{eq:CNN-equation}.
The CNN hyperparameters that are optimised during nested cross-validation are listed in Table \ref{table:class-hyper-parameter}. 
%and visually explained in Figure \ref{fig:CNN-fig}. 

\section{Classifier Training and Hyperparameter Optimisation}\label{sec:training}

\subsection{Nested cross-validation}\label{sec:cross-validation}
\begin{figure}
	\centerline{\includegraphics[width=0.8\textwidth]{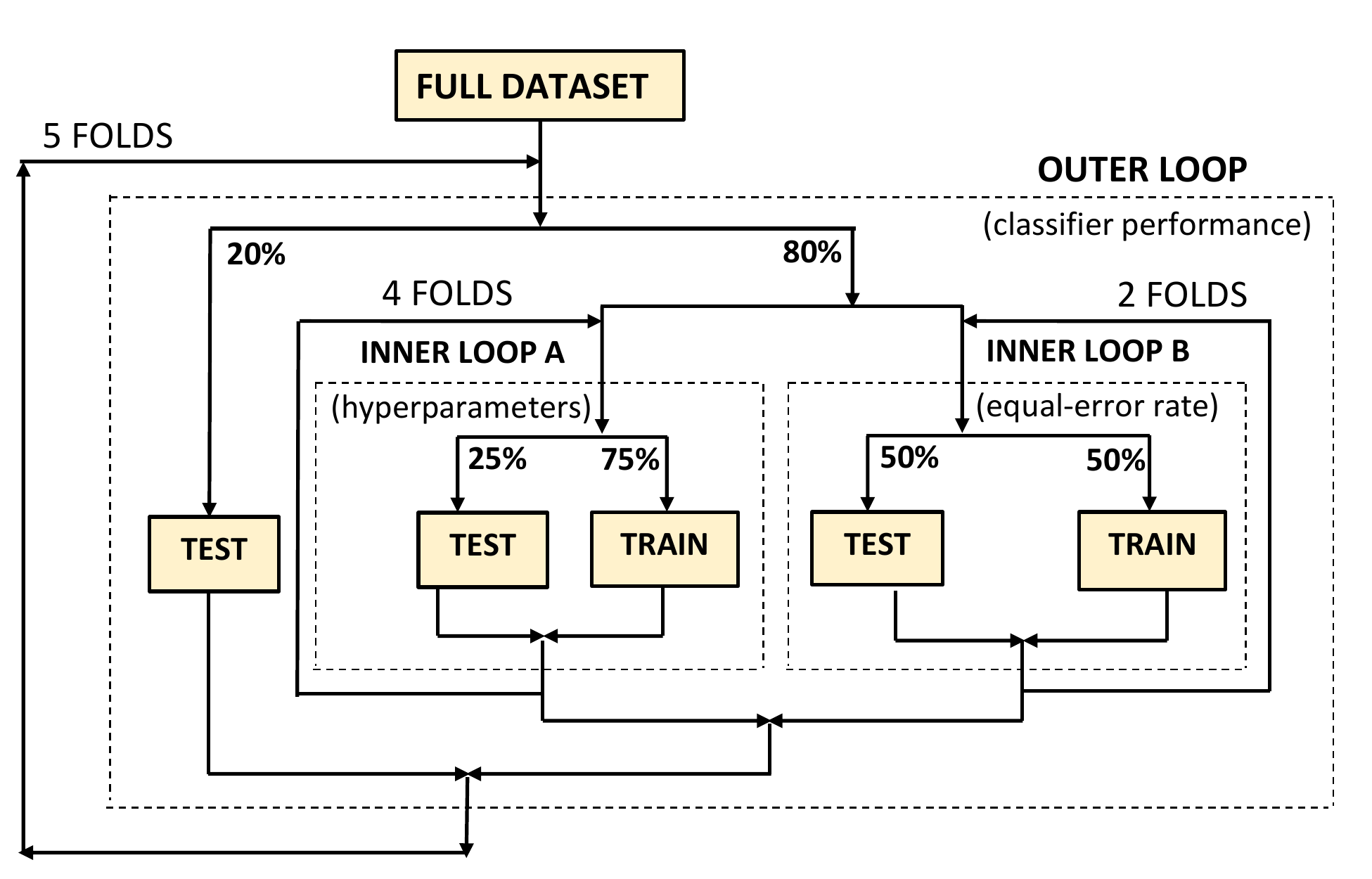}}
	\caption{\textbf{Nested k-fold cross-validation} was used for hyperparameter optimisation as well as the training and evaluation of all classifiers.}
	\label{fig:nested-k-fold}
\end{figure}

Because our dataset is small, we consistently used nested 5-fold cross-validation in all experiments. 
As shown in Figure~\ref{fig:nested-k-fold}, an outer loop divides the dataset into training (80\%) and testing (20\%) partitions where it is ensured that there is no patient overlap.
Within this outer loop, the training portion is again divided into two independent inner loops: one performing 4-fold and the other 2-fold cross-validation.
The former is used to optimise the hyperparameter listed in Table~\ref{table:class-hyper-parameter}, while the latter is used to determine the equal error rate which is used as part of the classifier decision. 
There was no patient overlap also within the inner loops and the gender balance was even. 
%The range of hyperparameters considered is shown in Table~\ref{table:class-hyper-parameter}. 

This cross-validation strategy makes the best use of our small dataset by allowing all patients to be used for training, hyperparameter optimisation, and final testing while ensuring unbiased optimisation and a strict per-patient separation between all training, development and testing portions while all folds contain the same proportion of both classes. 
%We use frames as training samples, and perform testing on a per-patient basis using cross-validation.

\subsection{Hyperparameter Optimisation}

Each classifier has different hyperparameters, as shown in Table~\ref{table:class-hyper-parameter}, which were optimised during nested k-fold cross-validation.
For the LR classifier, these were $\nu_1$, the gradient descent weight regularisation as well as $\nu_2$ and $\nu_3$, the lasso ($l1$) and ridge ($l2$) penalty estimators.
For the KNN classifier, the number of neighbours $\varkappa_1$ and the leaf-size $\varkappa_2$ were considered as hyperparameters, while for the SVM the regularisation strength $\zeta_1$ and the coefficient of the radial basis function kernel $\zeta_2$ were optimised.
For the MLP, the number of hidden neurons $\xi_1$, the $l2$ penalty $\xi_2$ and the stochastic gradient descent learning rate $\xi_3$ were considered.
Finally, for the CNN, the number of convolutional layers $\alpha_1$, the dropout rate $\alpha_2$ and the batch size $\alpha_3$ were optimised.

\subsection{Classifier Evaluation}

The area under the receiver operating characteristic (ROC) curve (AUC) has been used
as the primary evaluation metric of the classifier’s performance due to its higher degree of discriminancy for an imbalanced dataset such as ours and because it is widely used in medical diagnosis since 1970s \cite{rakotomamonjy2004optimizing,huang2005using,fawcett2006introduction}.
% as it is more discriminant
%
% capable to address the class imbalance present in a dataset such as ours~\cite{rakotomamonjy2004optimizing,fawcett2006introduction,huang2005using}. 
It also indicates how well the classifier has performed over a range of decision thresholds. 

\begin{table}[h]
	\caption{\textbf{Classifier hyperparameters} which are optimised using nested k-fold cross-validation.} % title name of the table
	\centering % centering table
	\begin{center}
		\begin{tabular}{ c | c | c  }
			\hline
			\hline
			\textbf{Hyperparameter} & \textbf{Classifier} & \textbf{Range} \\
			\hline
			\hline
			Regularisation Strength ($\nu_1$) & LR  & $10^{-7}$ to $10^{7}$ \\
			\hline
			$l1$ penalty ($\nu_2$) & LR   & 0 to 1 in steps of 0.05 \\
			\hline
			$l2$ penalty ($\nu_3$) & LR   & 0 to 1 in steps of 0.05 \\
			
			\hline
			No. of Neighbours ($\varkappa_1$) & KNN & 10 to 100 in steps of 10 \\
			\hline
			Leaf size ($\varkappa_2$) & KNN & 5 to 30 in steps of 5 \\
			
			\hline
			Regularisation Strength ($\zeta_1$) & SVM & $10^{-7}$ to $10^{7}$ \\
			\hline
			RBF kernel coefficient ($\zeta_2$) & SVM & $10^{-7}$ to $10^{7}$ \\
			
			\hline
			No. of hidden neurons ($\xi_1$) & MLP   & 10 to 100 in steps of 10 \\
			\hline
			$l2$ penalty ($\xi_2$) & MLP   & $10^{-7}$ to $10^{5}$ \\
			\hline
			Stochastic gradient descent learning rate ($\xi_3$) & MLP   & 0 to 1 in steps of 0.05 \\
			
			\hline
			No. of Convolutional Layers ($\alpha_1$) & CNN & $2^{k}$ where $k$ = 4, 5, 6 \\
			\hline
			Dropout Rate ($\alpha_2$) & CNN & 0.1 to 0.5 in steps 0.2 \\
			\hline
			Batch Size ($\alpha_3$) & CNN & $2^{k}$ where $k$ = 6, 7, 8 \\
			\hline
			\hline
		\end{tabular}
	\end{center}
	\label{table:class-hyper-parameter}
\end{table}

Receiver operating characteristic (ROC) curves were calculated within the inner loop of nested cross-validation. 
From these ROC curves, the decision threshold that achieves an equal-error-rate ($\gamma_{EE}$) was computed. 
%This is the threshold for which the difference between the classifier's true positive rate (TPR) and false positive rate (FPR) is minimised. 
%This $\gamma_{EE}$ was later used as an additional metric to evaluate the classifier's performance. 
%This nested cross-validation procedure makes best use of our small dataset by allowing all patients to be used for both training and testing purposes while ensuring unbiased hyper-parameter optimisation and a strict per-patient separation between cross-validation folds.
If the mean per-frame TB probability for a cough is $\hat{P}$ and the number of frames in a cough is $K$, then the cough is labelled as a TB cough, 
when $\hat{P} > \gamma_{EE}$, where: 

\begin{equation}
	\hat{P} = \frac{\sum\limits_{i=1}^{K} P(Y = 1|X, \theta)}{K} 
	\label{eq:P-hat}
\end{equation} 

Defining the indicator variable $C$ as:

\begin{equation}
	C = 
	\begin{cases}
		1 & \text{if } \hat{P}\geq \gamma_{EE}\\
		0              & \text{otherwise}
	\end{cases}
	\label{eq:C-sum}
\end{equation}

We define two TB index scores: $\text{TBI}_1$ and $\text{TBI}_2$ by following Equation \ref{eq:TBI-1} and \ref{eq:TBI-2}.  

\begin{equation}
	\text{TBI}_1 = \frac{\sum\limits_{i=1}^{N_1} C}{N_1} 
	\label{eq:TBI-1}
\end{equation}

\begin{equation}
	\text{TBI}_2 = \frac{\sum\limits_{i=1}^{N_2} P(Y = 1|X)}{N_2} 
	\label{eq:TBI-2}
\end{equation}

In Equation \ref{eq:TBI-1}, $N_1$ is the number of coughs obtained from the patient, while
in Equation~\ref{eq:TBI-2}, $N_2$ indicates the total number of frames of cough audio gathered from the patient.
Hence, Equation~\ref{eq:TBI-1} computes a per-cough average probability while Equation~\ref{eq:TBI-2} computes a per-frame average probability. 
Finally, a patient is classified as having TB when either the per-cough average probability is greater than 0.5, i.e. more than half of all coughs were classified as TB, or the per-frame average probability over all coughs is greater than $\gamma$. 

\begin{equation}
	\text{TB} = 
	\begin{cases}
		1 & \text{if } \text{TBI}_1 > 0.5 \\
		1 & \text{if } \text{TBI}_2 > \gamma \\
		0              & \text{otherwise}
	\end{cases}
	\label{eq:TB-final}
\end{equation}
%where $\gamma$ is varied along the ROC curve. 
Several variations of Equation~\ref{eq:TB-final} were considered, for example using only $\text{TBI}_1$ or only $\text{TBI}_2$ or including a threshold also for $\text{TBI}_1$. 
However, the presented formulation was found to be the most effective. 

Accuracies, positive predictive values (PPV) and negative predictive values (NPV) have also been calculated at the outer loop of the cross-validation scheme using Equation \ref{eq:accuracy}, \ref{eq:ppv} and \ref{eq:npv} respectively. 

\begin{equation}
	\text{Accuracy} = \frac{\text{TP} + \text{TN}}{\text{TP} + \text{FP} + \text{TN} + \text{FN}}
	\label{eq:accuracy}
\end{equation}

\begin{equation}
	\text{PPV} = \frac{\text{TP}}{\text{TP} + \text{FP}}
	\label{eq:ppv}
\end{equation}

\begin{equation}
	\text{NPV} = \frac{\text{TN}}{\text{FN} + \text{TN}}
	\label{eq:npv}
\end{equation}

Here, TP = true positives; TN = true negatives; FP = false positives and FN = false negatives.

%ROC-AUC has been calculated by following the Equation \ref{eq:AUC}, where $P$ = all positives, $N$ = all negatives, TPR = $\frac{\text{TP}}{P}$, FPR = $\frac{\text{FP}}{N}$, and $x$ is FPR. 
%
%\begin{equation}
%	\text{ROC-AUC} = \int_{0}^{1} \text{TPR} ({\text{FPR}}^{-1} (x)) \,dx 
%	\label{eq:AUC}
%\end{equation}

\section{Results}\label{sec:results}

\subsection{Classifier Performance}

The performance of classifiers trained and evaluated using the nested cross-validation procedure, described in Section~\ref{sec:cross-validation}, is shown in Table~\ref{table:classifier-results}.
The mean and standard deviation of the AUC was calculated over the outer cross-validation folds. 
The feature-extraction hyperparameters (Table~\ref{table:feat-hyper-parameter}) as well as the associated optimal classifier hyperparameters determined within the inner loop of nested cross-validation (Table~\ref{table:class-hyper-parameter}) are also given for each classifier architecture. 
Classifier hyperparameters producing the highest AUC in the outer folds have been noted as the `optimal classifier hyperparameters' in Table \ref{table:classifier-results}. 
Even for our small dataset, this procedure was computationally intensive.

\begin{table*}[h]
	\footnotesize
	\setlength{\tabcolsep}{4pt} % Default value: 6pt
	\caption{\textbf{Classifier performance in discriminating TB patients from non-TB patients:} The performance achieved by each considered classifier architecture in terms of the area under the ROC curve (AUC). Optimal classifier hyperparameters, determined during cross-validation, are also shown. Mean and standard deviations of AUC are calculated over the five outer folds of the nested k-fold cross-validation, shown in Figure \ref{fig:nested-k-fold}. } % title name of the table
	\centering % centering table
	\begin{center}
		\begin{tabular}{ c | c | c | c | c | c | c | c }
			\hline
			\hline
			\multirow{2}{*}{\textbf{Classifier}} & \textbf{Feature} & \textbf{Mean} & \textbf{Mean} & \textbf{Mean} & \textbf{Mean} & \textbf{Mean} & \textbf{Optimal Classifier} \\
			& \textbf{Hyperparameters} & \textbf{AUC} & \textbf{SD} & \textbf{Accuracy} & \textbf{PPV} & \textbf{NPV} & \textbf{Hyperparameters} \\
			\hline
			
			\hline
			\multirow{2}{*}{LR} & $\mathcal{M} = 13$; $\mathcal{F} = 2^{11}$; & \multirow{2}{*}{0.8000} &  \multirow{2}{*}{0.0519} & \multirow{2}{*}{78.82\%} & \multirow{2}{*}{75.41\%} & \multirow{2}{*}{83.29\%} & $\nu_1$=100, \\
			& $\mathcal{S} = 1$ &  &  &  &  &  & $\nu_2$=0.15, $\nu_3$=0.85 \\
			
			\hline
			\multirow{2}{*}{LR} & $\mathcal{M} = 13$; $\mathcal{F} = 2^{10}$; & \multirow{2}{*}{0.7842} & \multirow{2}{*}{0.0645} & \multirow{2}{*}{75.93\%} & \multirow{2}{*}{72.59\%} & \multirow{2}{*}{80.43\%} & $\nu_1$=0.001, \\
			& $\mathcal{S} = 1$ &  &  &  &  &  & $\nu_2$=0.25, $\nu_3$=0.75 \\
			
			\hline
			\multirow{2}{*}{LR} & $\mathcal{M} = 13$; $\mathcal{F} = 2^{11}$; & \multirow{2}{*}{0.7477} &  \multirow{2}{*}{0.0402} & \multirow{2}{*}{72.67\%} & \multirow{2}{*}{70.36\%} & \multirow{2}{*}{75.56\%} & $\nu_1$=10, \\
			& $\mathcal{S} = 1$ &  &  &  &  &  & $\nu_2$=0.15, $\nu_3$=0.85 \\
			
			\hline
			\multirow{2}{*}{\textit{LR}} & \textit{$\mathcal{M} = 26$; $\mathcal{F} = 2^{11}$;} & \multirow{2}{*}{\textit{0.8632}} & \multirow{2}{*}{\textit{0.0601}} & \multirow{2}{*}{\textit{84.54\%}} & \multirow{2}{*}{\textit{80.56\%}} & \multirow{2}{*}{\textit{89.71\%}} & \textit{$\nu_1$=0.01,} \\
			& \textit{$\mathcal{S} = 1$} &  &  &  &  &  & \textit{$\nu_2$=0.3, $\nu_3$=0.7} \\
			
			\hline
			\multirow{2}{*}{LR} & $\mathcal{M} = 26$; $\mathcal{F} = 2^{11}$; & \multirow{2}{*}{0.7845} &  \multirow{2}{*}{0.0491} & \multirow{2}{*}{76.22\%} & \multirow{2}{*}{72.9\%} & \multirow{2}{*}{80.67\%} & $\nu_1$=0.00001, \\
			& $\mathcal{S} = 1$ &  &  &  &  &  & $\nu_2$=0.45, $\nu_3$=0.55 \\
			
			\hline
			\multirow{2}{*}{LR} & $\mathcal{M} = 39$; $\mathcal{F} = 2^{11}$; & \multirow{2}{*}{0.7458} &  \multirow{2}{*}{0.0531} & \multirow{2}{*}{72.11\%} & \multirow{2}{*}{70.14\%} & \multirow{2}{*}{74.51\%} & $\nu_1$=0.0001, \\
			& $\mathcal{S} = 1$ &  &  &  &  &  & $\nu_2$=0.7, $\nu_3$=0.3  \\
			
			\hline
			\multirow{2}{*}{LR} & $\mathcal{M} = 39$; $\mathcal{F} = 2^{11}$; & \multirow{2}{*}{0.7402} &  \multirow{2}{*}{0.0455} & \multirow{2}{*}{73.02\%} & \multirow{2}{*}{70.57\%} & \multirow{2}{*}{76.14\%} & $\nu_1$=0.0001, \\
			& $\mathcal{S} = 1$ &  &  &  &  &  & $\nu_2$=0.4, $\nu_3$=0.6  \\
			
			\hline
			\multirow{2}{*}{LR} & $\mathcal{B} = 60$; $\mathcal{F} = 2^{11}$; & \multirow{2}{*}{0.7526} &  \multirow{2}{*}{0.0507} & \multirow{2}{*}{72.89\%} & \multirow{2}{*}{70.26\%} & \multirow{2}{*}{76.31\%} & $\nu_1$=0.1, \\
			& $\mathcal{S} = 1$ &  &  &  &  &  & $\nu_2$=0.45, $\nu_3$=0.55 \\

			\hline
			\multirow{2}{*}{KNN} & $\mathcal{M} = 26$; $\mathcal{F} = 2^{11}$; & \multirow{2}{*}{0.7701} & \multirow{2}{*}{0.0505} & \multirow{2}{*}{75.09\%} & \multirow{2}{*}{71.89\%} & \multirow{2}{*}{79.38\%} & $\varkappa_1$=80 \\
			& $\mathcal{S} = 1$ &  &  &  &  &  & $\varkappa_2$=20 \\
			
			\hline
			\multirow{2}{*}{KNN} & $\mathcal{M} = 26$; $\mathcal{F} = 2^{11}$; & \multirow{2}{*}{0.7394} & \multirow{2}{*}{0.0385} & \multirow{2}{*}{70.76\%} & \multirow{2}{*}{68.95\%} & \multirow{2}{*}{72.96\%} & $\varkappa_1$=60 \\
			& $\mathcal{S} = 1$ &  &  &  &  &  & $\varkappa_2$=15 \\

			\hline
			\multirow{2}{*}{SVM} & $\mathcal{M} = 26$; $\mathcal{F} = 2^{11}$; & \multirow{2}{*}{0.7435} & \multirow{2}{*}{0.0543} & \multirow{2}{*}{71.91\%} & \multirow{2}{*}{69.69\%} & \multirow{2}{*}{74.7\%} & $\zeta_1$=0.01 \\
			& $\mathcal{S} = 1$ &  &  &  &  &  & $\zeta_2$=100 \\
			
			\hline
			\multirow{2}{*}{SVM} & $\mathcal{M} = 39$; $\mathcal{F} = 2^{10}$; & \multirow{2}{*}{0.7291} & \multirow{2}{*}{0.0495} & \multirow{2}{*}{70.05\%} & \multirow{2}{*}{67.95\%} & \multirow{2}{*}{72.71\%} & $\zeta_1$=0.001 \\
			& $\mathcal{S} = 1$ &  &  &  &  &  & $\zeta_2$=0.0001 \\

			\hline
			\multirow{2}{*}{MLP} & $\mathcal{M} = 13$; $\mathcal{F} = 2^{11}$; & \multirow{2}{*}{0.7389} & \multirow{2}{*}{0.0457} & \multirow{2}{*}{71.16\%} & \multirow{2}{*}{68.91\%} & \multirow{2}{*}{74.02\%} & $\xi_1$=80, \\
			& $\mathcal{S} = 1$ &  &  &  &  &  & $\xi_2$=0.0001, $\xi_3$=0.65 \\
			
			\hline
			\multirow{2}{*}{MLP} & $\mathcal{M} = 26$; $\mathcal{F} = 2^{11}$; & \multirow{2}{*}{0.8000} & \multirow{2}{*}{0.0391} & \multirow{2}{*}{77.87\%} & \multirow{2}{*}{75.13\%} & \multirow{2}{*}{81.28\%} & $\xi_1$=50, \\
			& $\mathcal{S} = 1$ &  &  &  &  &  & $\xi_2$=0.001, $\xi_3$=0.55 \\
			
			\hline
			\multirow{2}{*}{MLP} & $\mathcal{M} = 39$; $\mathcal{F} = 2^{11}$; & \multirow{2}{*}{0.7742} & \multirow{2}{*}{0.0409} & \multirow{2}{*}{76.47\%} & \multirow{2}{*}{73.63\%} & \multirow{2}{*}{80.09\%} & $\xi_1$=30, \\
			& $\mathcal{S} = 1$ &  &  &  &  &  & $\xi_2$=0.01, $\xi_3$=0.35 \\

			\hline
			\multirow{2}{*}{CNN} & $\mathcal{M} = 26$; $\mathcal{F} = 2^{11}$; & \multirow{2}{*}{0.7109} & \multirow{2}{*}{0.0409} & \multirow{2}{*}{68.89\%} & \multirow{2}{*}{67.7\%} & \multirow{2}{*}{70.25\%} & $\alpha_1$=32 \\
			& $\mathcal{S} = 1$ &  &  &  &  &  & $\alpha_2$=0.3, $\alpha_3$=128 \\
			
			\hline
			\multirow{2}{*}{CNN} & $\mathcal{M} = 39$; $\mathcal{F} = 2^{10}$; & \multirow{2}{*}{0.7001} & \multirow{2}{*}{0.0301} & \multirow{2}{*}{68.71\%} & \multirow{2}{*}{67.52\%} & \multirow{2}{*}{70.07\%} & $\alpha_1$=64 \\
			& $\mathcal{S} = 1$ &  &  &  &  &  & $\alpha_2$=0.1, $\alpha_3$=128 \\
			
			\hline
			\hline
		\end{tabular}
	\end{center}
	\label{table:classifier-results}
\end{table*}

\subsection{Feature Selection}
\begin{figure}[h!]
	\centerline{\includegraphics[width=0.9\textwidth]{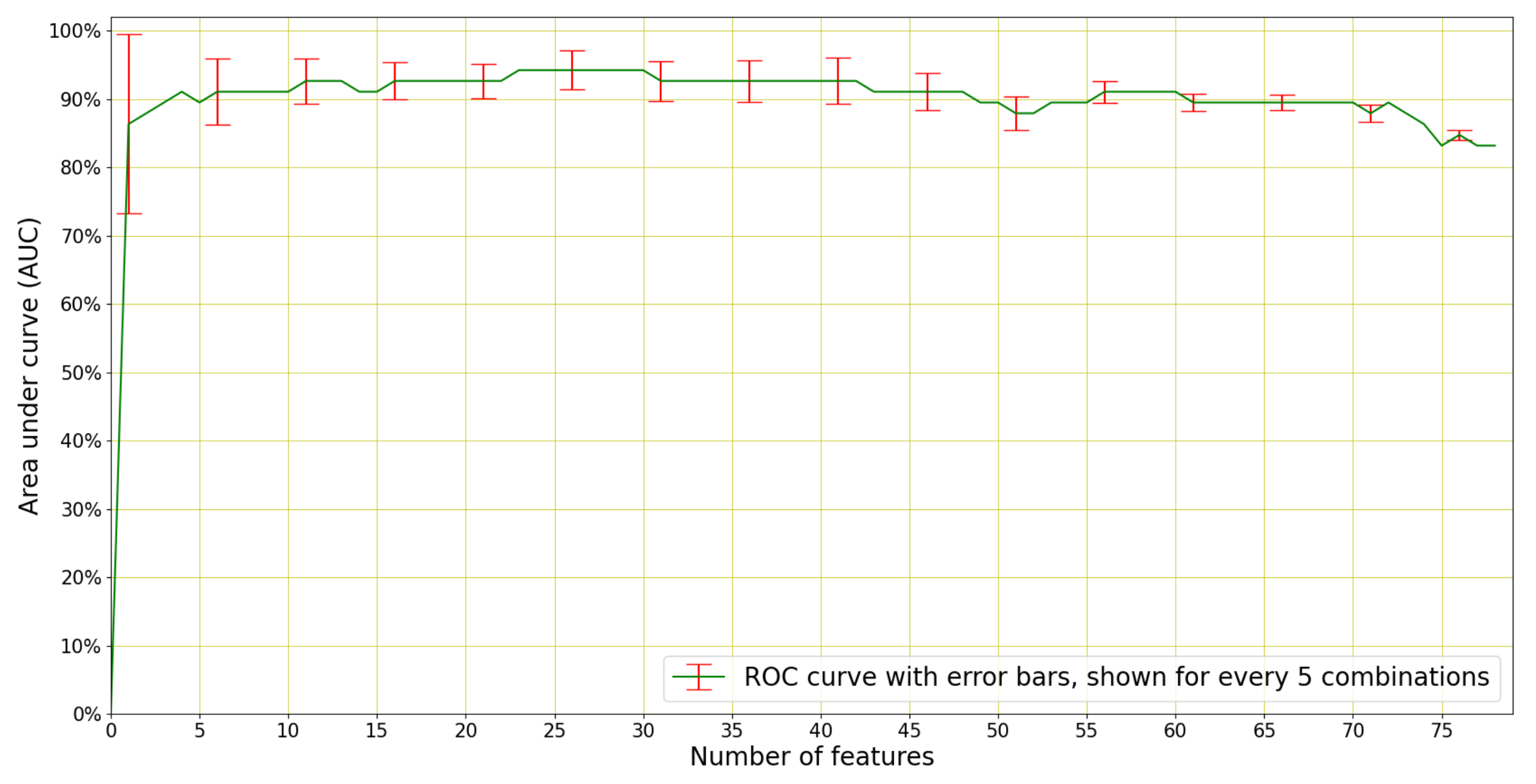}}
	\caption{\textbf{Sequential Forward Selection} (SFS) applied to the best performing system (LR) in Table~\ref{table:classifier-results} which uses a total 78 features (26 MFCCs with appended velocity and acceleration). A maximum AUC of 0.94 is achieved using the best 23 features to discriminate TB patients from non-TB patients. The error bars (standard deviation) of the AUC is indicated every 5 features.}
	\label{fig:best-SFS}
\end{figure}

As an additional experiment, sequential forward selection (SFS)~\cite{devijver1982pattern} was applied to discover the best performing individual features responsible for distinguishing between TB and non-TB coughs.
SFS is a greedy selection procedure that, starting from a single feature, sequentially finds the additional feature that contributes the most to classification performance. 
SFS was applied within the inner cross-validation loop to the best performing system in Table~\ref{table:classifier-results}, which uses 26 MFCCs with appended velocity ($\Delta$) and acceleration ($\Delta \Delta$) coefficients, and therefore 78 features in total. 
The results of this selection are shown in Figure \ref{fig:best-SFS}. 

We see in Figure \ref{fig:best-SFS} that optimal performance is achieved using 23 of the total 78 features and near-optimal performance is achieved using as few as four features. 
These best-four features are the $3^{rd}$, the $11^{th}$, the velocity ($\Delta$) of the $14^{th}$, and the $12^{th}$ MFCCs. 
It is interesting to note that the first acceleration ($\Delta \Delta$) feature to be chosen appears only in the $9^{th}$ position, after the best-four and the $5^{th}$, $17^{th}$, $7^{th}$, and $18^{th}$ MFCCs. 
Figure \ref{fig:best-SFS} also shows that there are several features that lead to deteriorated performance and should therefore be omitted from the classifier. 

\begin{figure}[h!]
	\centerline{\includegraphics[width=0.95\textwidth]{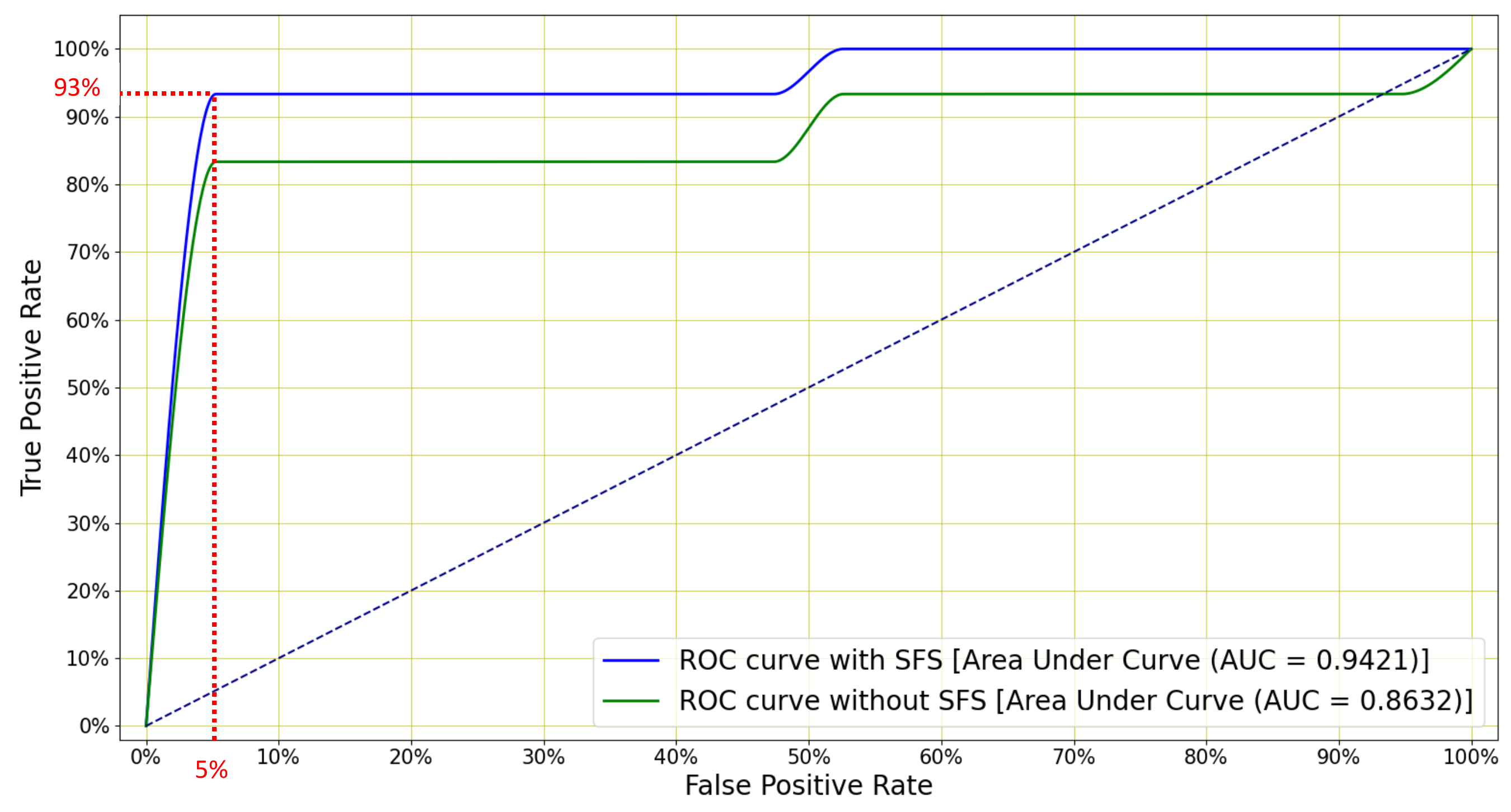}}
	\caption{\textbf{Mean ROC curves} for the LR classifier when distinguishing between TB and non-TB patients with and without SFS (Figure~\ref{fig:best-SFS}). The former uses all 78 features while the latter retains only the 23 best features. A sensitivity of 93\% is achieved at 95\% specificity and this exceeds the minimum WHO specification for community-based TB triage testing.}
	\label{fig:best-LR-result}
\end{figure}

Finally, Figure~\ref{fig:best-LR-result} shows the ROC curves for the classifier subjected to SFS, both when using all 78 and when using the best 23 features.
We see that SFS affords better classification performance across a wide range of operating conditions.

\section{Discussion}
\label{sec:discussion}

The results in Table~\ref{table:classifier-results} show that LR outperforms the other four classifiers, achieving an AUC of 0.86 and an accuracy of 84.54\%. %0.8632. 
Figure~\ref{fig:best-SFS} shows that, by applying SFS to this LR classifier and retaining the top 23 features, the AUC is further improved to 0.94. %  0.9421 
Among these 23 selected features, velocity and acceleration coefficients appear only once in the top nine features. 
Hence near-optimal performance can be achieved by using MFCCs without added velocity or acceleration. 
Figure~\ref{fig:best-LR-result} shows that this system is able to achieve a sensitivity of 93\% at a specificity of 95\% which exceeds the minimal requirement for a community-based TB triage test of 90\% sensitivity at 70\% specificity set by the WHO~\cite{world2014high}. 
%Table \ref{table:classifier-results} also shows that the LR has achieved the highest accuracy of 84.54\%. 

The LR model was intended to be our baseline, and we had hoped to improve classification performance using more complex models, such as the MLP and CNN which both contain multiple neurons and can model non-linear relationships.
However, it seems that our dataset is too small and perhaps too noisy for the greater flexibility of neural networks to be of benefit.
We note that other researchers have also reported the superiority of LR in some clinical prediction tasks~\cite{christodoulou2019systematic}. 
%Although we aim to collect more data, studies such as \cite{knocikova2008wavelet} present convincing results on even smaller dataset than ours. 
%Although a larger dataset might benefit the performance of deep neural networks such as the CNN, 
We also note that all our classifiers have been evaluated within the same nested k-fold cross-validation scheme. 
Hence, even though the more complex neural network architectures such as the CNN might in future benefit from more training data, our comparison between the classifiers is justifiable and the achieved performance is a reflection of what is currently possible. 

Table~\ref{table:classifier-results} also shows that, for all classifiers, splitting the cough audio signal into multiple sections does not lead to improved performance, 
which suggests that the acoustic information in all phases of a cough is equally important for the purposes of TB classification. 
However, increasing the number of MFCCs does provide consistent improvements, as does the use of longer frames. 
MFCCs are shown to outperform linearly spaced log-filterbank energies as features, which is in contrast to the indications in our previous work~\cite{botha2018detection}.
However, in our previous work, only a classical MFCC configuration (13 coefficients, with appended velocity and acceleration) was considered.
Here, we find that better performance can be achieved using a larger number of MFCCs than is necessary to model the discriminatory characteristics of human hearing.
This again leads us to conclude that the classifier is to some extent basing its decision on acoustic information not perceivable by a human listener. 
%This makes TB cough classification task more challenging as unlike croup coughs, which can be classified with higher accuracy \cite{sharan2017cough,sharan2018automatic}, informal listening suggests that TB coughs do not appear different to the human ear \cite{botha2018detection}. 

When compared with the simpler log-filterbank energies, MFCCs have the additional advantage of providing a simple and effective way to compensate for convolutional channel variability by means of mean normalisation. 
The dataset we use in this work is more noisy and less controlled than the one we used in our previous work. 
For example, the microphone position is generally a little different between recordings since it does not remain in the collection booth overnight.
Hence the advantages of channel normalisation may weigh more strongly for the dataset we consider here.

Finally, we have seen that the number of features can be reduced in a greedy fashion to optimise performance.
The highest AUC is achieved when using 23 of the possible 78 features, and near-optimal performance can be achieved using as few as four features.
This is of particular importance with a view to implement audio TB screening on mobile computing devices, such as smartphones, since computational effort is saved by reducing the dimensionality of the feature vector. 
Implementation on a consumer mobile device would make the algorithm portable, inexpensive and easy to apply, which makes it attractive in under-resourced environments.

\section{Conclusion and Future Work}\label{sec:conclusion}

We have shown for the first time that it is possible to automatically distinguish between the forced coughing sounds of tuberculosis (TB) patients and the coughing sounds of patients with other lung ailments. 
This strengthens our previous work which indicated that such discrimination is possible between the coughs of TB patients and healthy controls. 
In contrast to diseases such as croup coughs, which can be classified with higher accuracy \cite{sharan2017cough,sharan2018automatic}, the sounds of coughs by TB sufferers do not appear to possess obviously identifiable characteristics. 
This view is based on informal listening tests of our data, as well as the personal opinions of several medical practitioners who we consulted during the course of this research. 
The identification of precisely which aspects of the cough signal are important for TB classification are a subject of our ongoing work.

Our experiments are based on a newly-compiled dataset recorded in a noisy primary healthcare clinic. 
Hence, we also show that TB cough classification is possible in the type of real-world environment that may be expected at a screening facility in a developing country. 
Using nested cross-validation, five machine learning classifiers were evaluated.
By applying logistic regression (LR) and performing sequential forward selection (SFS) to select the top 23 of 78 high-resolution MFCC features, an area under the ROC curve (AUC) of 0.94 was achieved which shows that MFCCs without velocity or acceleration can produce near-optimal performance. 
This classifier achieves 93\% sensitivity at 95\% specificity, which exceeds the 90\% sensitivity at 70\% specificity specification considered by the World Health Organisation (WHO) as a minimal requirement for community-based triage testing~\cite{world2014high}.

The proposed screening by automatic analysis of coughing sounds is non-intrusive, can be applied without specialist medical expertise or laboratory facilities, produces a result quickly and can be implemented on readily-available and inexpensive consumer hardware, such as a smartphone.
It therefore may represent a useful tool in the fight against TB especially in developing countries where the TB burden is high, such as our own setting in Cape Town, South Africa \cite{blaser2016tuberculosis,mulongeni2019hiv}.
Recent studies have shown that, in South Africa, there are currently on average between 600 and 700 TB cases per 100,000 people 
\cite{global-TB-report,nicd-TB-report,TB-facts}. 
%\cite{nicd-TB-report}
%\cite{TB-facts}. 

Our study has several limitations and we aim to improve those in our future work. 
Firstly, although our dataset is unique, it is also rather small compared to some other datasets used for cough detection \cite{pahar2021deep} and classification \cite{sharma2020coswara}.
We believe this is why more advanced classifiers, such as a convolutional neural network (CNN), did not offer any performance advantage in our experiments \cite{pahar2021covidbreath}.
We are extending the dataset, hoping that this will allow such more advanced classifiers to perform better than the LR baseline. %\cite{ahmed2019deeplung,wang2015audio}. 
Secondly, we are currently using only the recordings of the cough sounds as a basis of classification.
The speech audio which was also recorded as part of our data collection might allow classifier accuracy to be improved and this investigation is currently ongoing. 
Thirdly, the manually annotated cough events sometimes contain multiple bursts of cough onsets and methods that identify such bursts within a cough event automatically are also a subject of our ongoing investigation. 
Fourthly, we have evaluated our classifier using only a single dataset.
To better establish the ability of the classifier to correctly process truly unseen data, an additional validation-only dataset is required.
We are in the process of planning such a data collection effort, where recordings will be made in different but also noisy primary healthcare environments. 
Fifthly, participants were recruited into this study based on a self-reported cough.
This approach may miss patients who have a cough but do not report it. To address this, further studies without cough as an eligibility criterion are required. 
Finally, the proposed system is not yet ready for practical implementation.
The automatic detection of the cough within the recorded audio must be considered, as well as the practical integration of our classifier on a mobile device, 
as well as the consideration of additional audio captured by a stethoscope \cite{pasterkamp1997respiratory}, also form part of our ongoing work.

\section{Acknowledgements}

This project was partially supported by the South African Medical Research Council (SAMRC) through its Division of Research Capacity Development under the SAMRC Intramural Postdoctoral programme from funding received from the South African National Treasury. 
We also acknowledge
%, as well as an EDCTP2 programme supported by the European Union 
%%(TMA2017CDF-1885). 
%(grant SF1401, OPTIMAL DIAGNOSIS; RIA2020I-3305, CAGE-TB).
funding from the EDCTP2 programme supported by the European Union (grant SF1401, OPTIMAL DIAGNOSIS; grant RIA2020I-3305, CAGE-TB) and the National Institute of Allergy and Infection Diseases of the National Institutes of Health (U01AI152087).

We would like to thank the South African Centre for High Performance Computing (CHPC) for providing computational resources on their Lengau cluster for this research and gratefully acknowledge the support of Telkom South Africa. 
We also thank the Clinical Mycobacteriology \& Epidemiology (CLIME) clinic team for assisting in data collection, especially Sister Jane Fortuin and Ms. Zintle Ntwana.

The content and findings reported are the sole deduction, view and responsibility of the researcher and do not reflect the official position and sentiments of the SAMRC, EDCTP2, European Union or the funders. 
The authors have confirmed that any identifiable participants in this study have given their consent for publication. 

\section{References}

\bibliographystyle{dcu}
\bibliography{reference}

\end{document}